\documentclass[12pt,preprint]{aastex}

\shorttitle{Transverse oscillations of two coronal loops}
\shortauthors{M. Luna et al.}
\begin{document}

\title{Transverse oscillations of two coronal loops}

\author{M. Luna\altaffilmark{1}, J.
Terradas\altaffilmark{1,2}, R. Oliver\altaffilmark{1}, and J.L.
Ballester\altaffilmark{1}}
%\email{manuel.luna@uib.es}
%\email{jaume.terradas@uib.es}
%\email{ramon.oliver@uib.es}
%\email{dfsjlb0@uib.es}

\altaffiltext{1}{Departament de F\'{\i}sica, Universitat de les Illes Balears,
07122 Palma de Mallorca, Spain. Email: manuel.luna@uib.es,
jaume.terradas@uib.es, ramon.oliver@uib.es and joseluis.ballester@uib.es}
\altaffiltext{2}{Centre for Plasma Astrophysics, Katholieke Universiteit Leuven,
Celestijnenlaan 200B, B-3001 Leuven, Belgium}

\begin{abstract}
We study transverse fast magnetohydrodynamic waves in a system of two coronal
loops modeled as smoothed, dense plasma cylinders in a uniform magnetic field.
The collective oscillatory properties of the system due to the interaction
between the individual loops are investigated from two points of view. Firstly,
the frequency and spatial structure of the normal modes are studied. The system
supports four trapped normal modes in which the loops move rigidly in the
transverse direction. The direction of the motions is either parallel or
perpendicular to the plane containing the axes of the loops. Two of these modes
correspond to oscillations of the loops in phase, while in the other two they
move in antiphase. Thus, these solutions are the generalization of the kink mode
of a single cylinder to the double cylinder case. Secondly, we analyze the
time-dependent problem of the excitation of the pair of tubes. We find that
depending on the shape and location of the initial disturbance, different normal
modes can be excited. The frequencies of normal modes are accurately recovered
from the numerical simulations. In some cases, because of the simultaneous
excitation of several eigenmodes, the system shows beating.
\end{abstract}


\section{Introduction}

Transverse coronal loop oscillations have been studied in recent years after
being observed for the first time by the Transition Region and Coronal Explorer
(TRACE) in 1998 \citep[see for example][]{aschwanden1999, aschwanden2002,
schrijver2002, verwichte2004}. These oscillations were initiated shortly after a
solar flare that disturbed the loops. Much before the TRACE observations, the
theory of loop oscillations was developed
\citep{spruit1981,edwin&roberts1983,cally1983} and the different kinds of
oscillations were studied. The observed transverse motions have been interpreted
in terms of the excitation of the fast magnetohydrodynamic (MHD) fundamental
kink mode.

Most analytical studies about transverse loop oscillations have only considered
the properties of individual loops, but in many cases loops belong to complex
active regions where they are usually not isolated. For example,
\citet{verwichte2004} reported complex transverse motions of loops in a
post-flare arcade. In particular, loops D and E \citep[see Fig.~1
of][]{verwichte2004} show bouncing displacements  with oscillations in phase and
antiphase that repeat in time. The same behavior of the movements in a loop
bundle can be observed in the event of March 23, 2000 of the compact flare
recorded by TRACE \citep[see][]{schrijver2002}. Additionally, antiphase
oscillations of adjacent loops have also been reported in
\citet{schrijver2000,schrijver2002}. These observations suggest that there are
interactions between neighboring loops and that the dynamics of loop systems is
not simply the sum of the dynamics of the individual loops.

On the other hand, it is currently debated whether active region coronal loops
are monolithic \citep{aschwanden2005} or multi-stranded
\citep{klimchuk2006,deforest2007}. The strands are considered as mini-loops for
which the heating and plasma properties are approximately uniform in the
transverse direction. In the multi-stranded model it is suggested that loops are
formed by bundles of several tens or several hundreds of physically related
strands \citep{klimchuk2006}. \citet{lopezfuentes2006} suggest that these
strands wrap around each other in complicated ways due to the random motion of
the foot points in the solar surface. These models explain the constant width
and symmetry of the loops as observed with current X-ray and EUV telescopes.

From the observations, it is thus necessary to study not only individual loops
but also how several loops or strands can oscillate as a whole, since their
joint dynamics can be different from those of a single loop. Little work has
been done on composite structures so far. \citet{berton1987} studied the
magnetohydrodynamic normal modes of a periodic magnetic medium, while other
authors, for example \citet{bogdan&fox1991, keppens1994}, analyzed the
scattering and absorption of acoustic waves by bundles of magnetic flux tubes
with sunspot properties. \citet{kris93, kris94} studied numerically the
propagation of fast waves in two slabs unbounded in the longitudinal direction.
On the other hand, in \citet{diaz2005} the oscillations of the prominence thread
structure were investigated. These authors found that in a system of equal
fibrils the only non-leaky mode is the symmetric one, with all fibrils
oscillating in spatial phase with the same frequency. Finally, \citet{luna}
found that in a system of  two coronal slabs, the symmetric and antisymmetric
modes can be trapped and that an initial disturbance can excite these modes,
which are readily detectable after a brief transient phase. If the fundamental
symmetric mode and the antisymmetric first harmonic are excited at the same
time, a beating phenomenon takes place. In such a case, the loops interchange
energy periodically. In any case, all these authors found that a system of
several loops behave differently from an individual loop.

Here we consider a more complex system than those studied in previous works. Our
model consists of two parallel cylinders, without gravity and curvature. This
model allows us to study the interaction between loops and the collective
behavior of the system. We study the normal modes and also solve the
time-dependent problem of the excitation of transverse coronal loop
oscillations.  We concentrate on a planar pulse excitation and compare the
results of the simulations with the eigenmodes of the configuration.

This paper is organized as follows. In \S \ref{loop_model} the loop model is
presented. In \S \ref{normal_modes_sec} the normal modes are calculated and  the
frequencies and spatial distribution of the eigenfunctions are studied. The
time-dependent problem is considered in \S \ref{time_dependent_analysis}, where
the velocity and pressure field distribution are analyzed for different
incidence angles of the initial perturbation. In \S \ref{2d_beating} the loop
motions are studied and the beating is analyzed. Finally, in \S
\ref{discussion_conclusions} the results are summarized and the main conclusions
are drawn.

\section{Equilibrium configuration and basic equations}
\label{loop_model}

\begin{figure}[!ht]
\center
\resizebox{8cm}{!}{\includegraphics{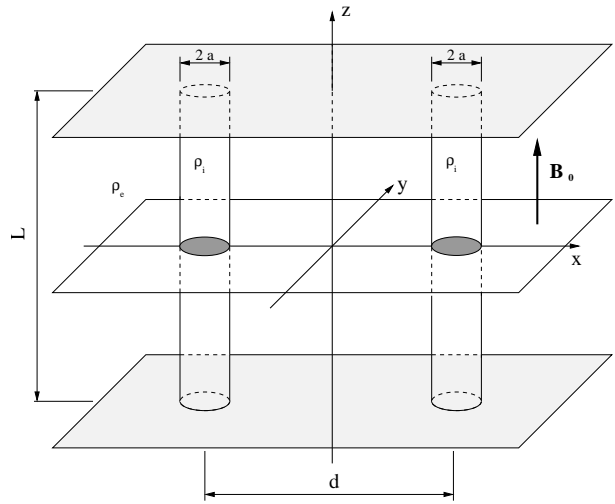}}
\caption{
\label{sketch}
\small Sketch of the model, that consists of two straight cylinders immersed in
the coronal medium. Hereafter the total pressure and the velocity fields are
plotted in the $xy$-plane, shown as a white slice.}
\end{figure}

The simplest way to investigate the interaction of a set of loops is to consider
a pair of loops in slab geometry. In \citet{luna} this model was studied in
detail using  the  ideal MHD equations and the zero-$\beta$ plasma limit. Here a
more realistic model is considered. The equilibrium configuration consists of a
system of two parallel homogeneous straight cylinders of radius $a$, length
$L$,  and separation between centers $d$ (see Fig.~~\ref{sketch}). We assume the
following equilibrium plasma density profile:
\begin{eqnarray*}
\rho_\mathrm{0} (x,y)=\left\{
\begin{array}{lll}
\rho_\mathrm{e}, & \;{\textrm{\normalsize if $r_\mathrm{1}> a$ and $r_\mathrm{2}
> a$},} \\
\rho_\mathrm{i}, & \;{\textrm{\normalsize if
 $r_\mathrm{1} \leq a$ or $r_\mathrm{2} \leq a$},}
 \end{array}
\right.
\end{eqnarray*}
where $x$, $y$ are the Cartesian coordinates and $r_\mathrm{1}$ and
$r_\mathrm{2}$, defined as $r_\mathrm{1}^2=(x+d/2)^2+y^2$ and
$r_\mathrm{2}^2=(x-d/2)^2+y^2$, are the distances from the point $(x,y)$ to the
centers of the left and right loops, respectively. In the previous expression 
$\rho_\mathrm{e}$ and $\rho_\mathrm{i}$ are the densities in the external medium
or corona and the loop ($\rho_\mathrm{i}>\rho_\mathrm{e}$), respectively.
Hereafter, we use a density contrast $\rho_\mathrm{i}/\rho_\mathrm{e}=10$. 
\clearpage
\clearpage
The loop centers lie on the $x$-axis at $x=d/2$ for the right loop and $x=-d/2$
for the left loop. The configuration is symmetric with respect to the $yz$-plane
and  the $z$-axis is parallel to the axes of the cylinders. The tubes and the
environment are permeated by a uniform magnetic field along the $z$-direction
($\mathbf{B} = B_\mathrm{0} \mathbf{e_\mathrm{z}}$). The Alfv\'en speed,
$v_\mathrm{A} = B_\mathrm{0}/\sqrt{\mu \rho}$, takes the value $v_\mathrm{A i}$
inside the loop and $v_\mathrm{Ae}$ in the surrounding corona ($v_\mathrm{A i} <
v_\mathrm{A e}$). 

Linear perturbations about this equilibrium for a perfectly conducting fluid in
the zero-$\beta$ limit can be readily described using the ideal MHD equations in
Cartesian coordinates. The velocity is denoted by $\mathbf{v}= \left(v_x, v_y, 0
\right)$ and $\mathbf{B}= \left(B_x, B_y, B_z \right)$ is the magnetic field
perturbation. We have assumed a $z$-dependence of the perturbations of the form
$e^{-i k_z z}$. In this model we consider the photosphere as two infinitely
dense planes located at $z=\pm L/2$. The loop feet are anchored in these planes
and so the fluid velocity is zero at these positions (this is the so-called
line-tying effect). This condition produces a quantization of the $z$-component
of the wave-vector to $k_z=n \pi/L$. Hereafter we concentrate on the fundamental
mode, with $n=1$. The total pressure perturbation is
\begin{equation}\label{total_pressure}
 p_\mathrm{T}=\frac{B_\mathrm{0}}{\mu} B_z,
\end{equation}
and coincides with the magnetic pressure perturbation in the zero-$\beta$ limit.

\section{Normal modes}
\label{normal_modes_sec}

Analytical solutions to the eigenvalue problem of the previous model (assuming 
a temporal dependence of the form $e^{i \omega t}$) are very difficult to derive
due to the geometry of the system. The methods used for a single cylinder
\citep[see][]{edwin&roberts1983} cannot be applied to the study of two tubes.
One way to solve the problem is to use scattering theory, see for example
\citet{edwin&roberts1983}, \citet{bogdan&knolker1990},
\citet{bogdan&zeibel1985}, \citet{bogdan&fox1991} and \citet{keppens1994}.
Another way is  to solve the eigenvalue problem given by the ideal MHD equations
numerically. We have used this approach and we have done the computations with 
the PDE2D code \citep{sewell}. We have used bicylindrical orthogonal
coordinates, defined by the transformation 
\begin{eqnarray} 
x=\frac{d/2 \sinh v}{\cosh v - \cos u} ~ ,~ y=\frac{d/2 \sin u}{\cosh v - \cos
u} ~ , 
\end{eqnarray} 
where $0 \le u < 2\pi$ and $-\infty < v < \infty$. The loop boundaries are
coordinate surfaces at $v=\pm \mathrm{arcsinh} ~\frac{d}{2 a}$, where the
positive and negative signs correspond to the right and left tubes,
respectively. We impose the restriction that the solutions tend to zero at large
distances from the cylinders, i.e. we seek trapped mode solutions. 

\begin{figure}[!ht]
\center
\mbox{\hspace{0.cm}\hspace{8.cm}\hspace{8.cm}}\vspace{-2.cm}
\includegraphics[width=7.5cm]{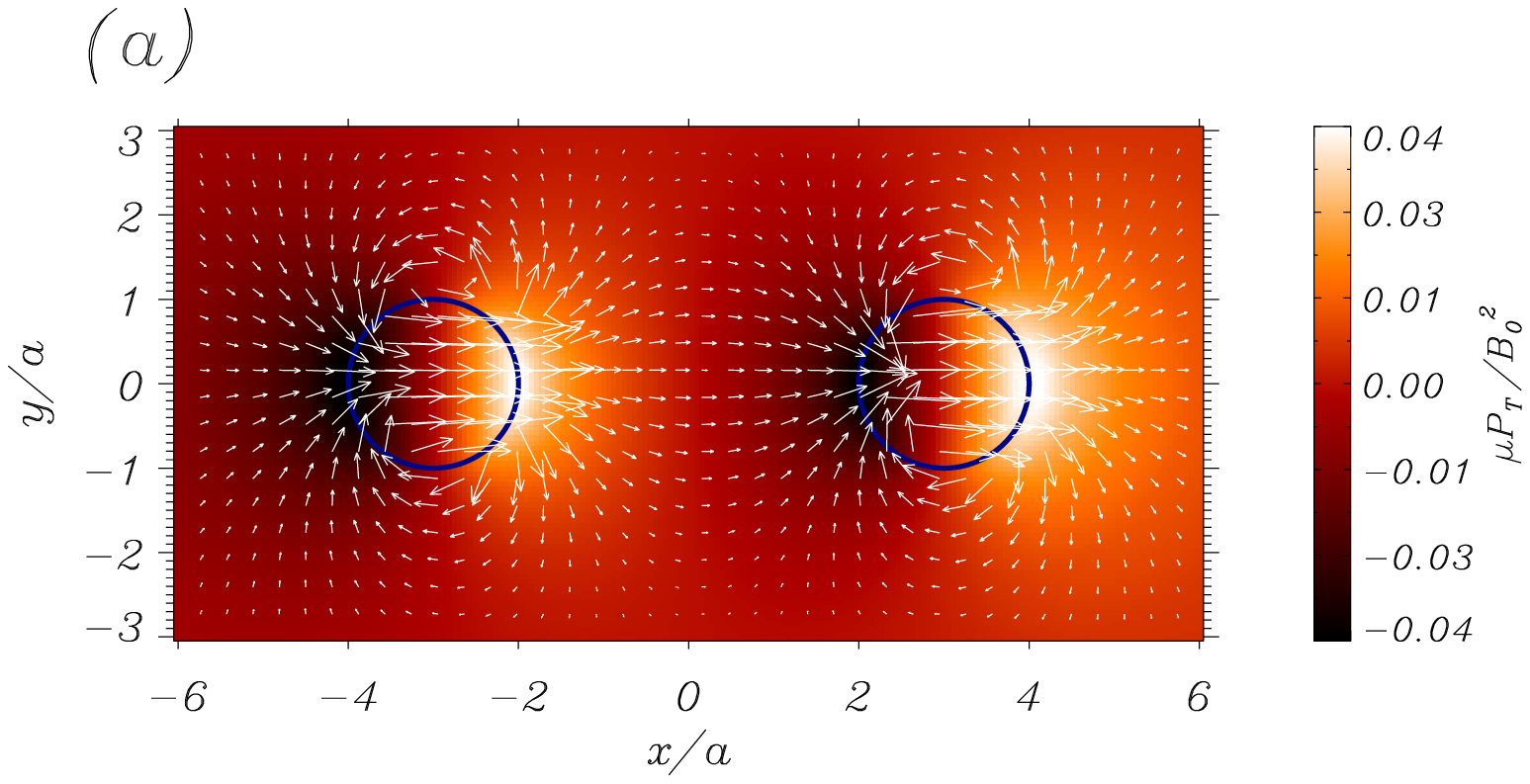}\hspace{0.8cm}\includegraphics[width=7.5cm]{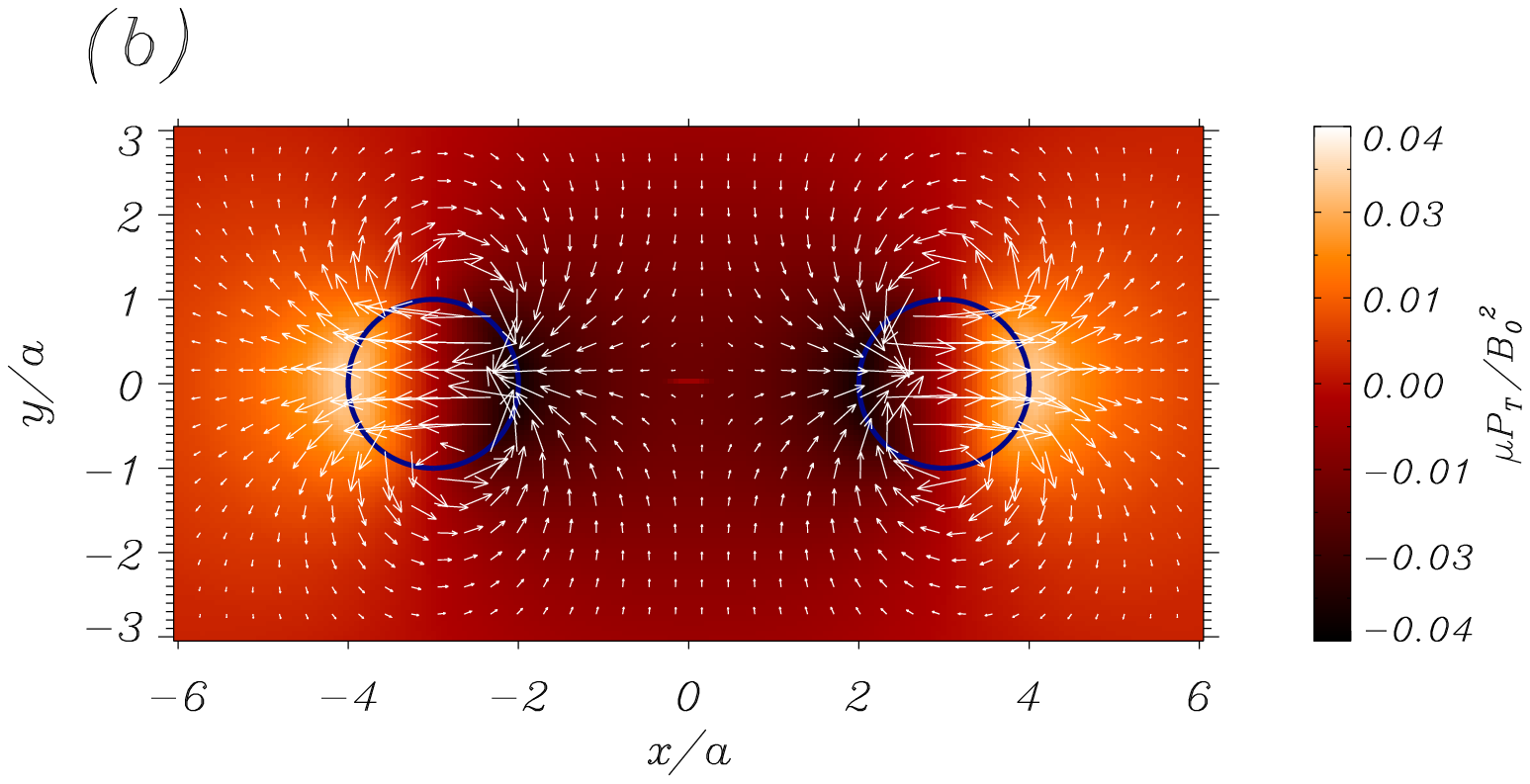}\vspace{-4.cm}
\mbox{\hspace{-0.cm}\hspace{8.cm}\hspace{8.cm}}
\includegraphics[width=7.5cm]{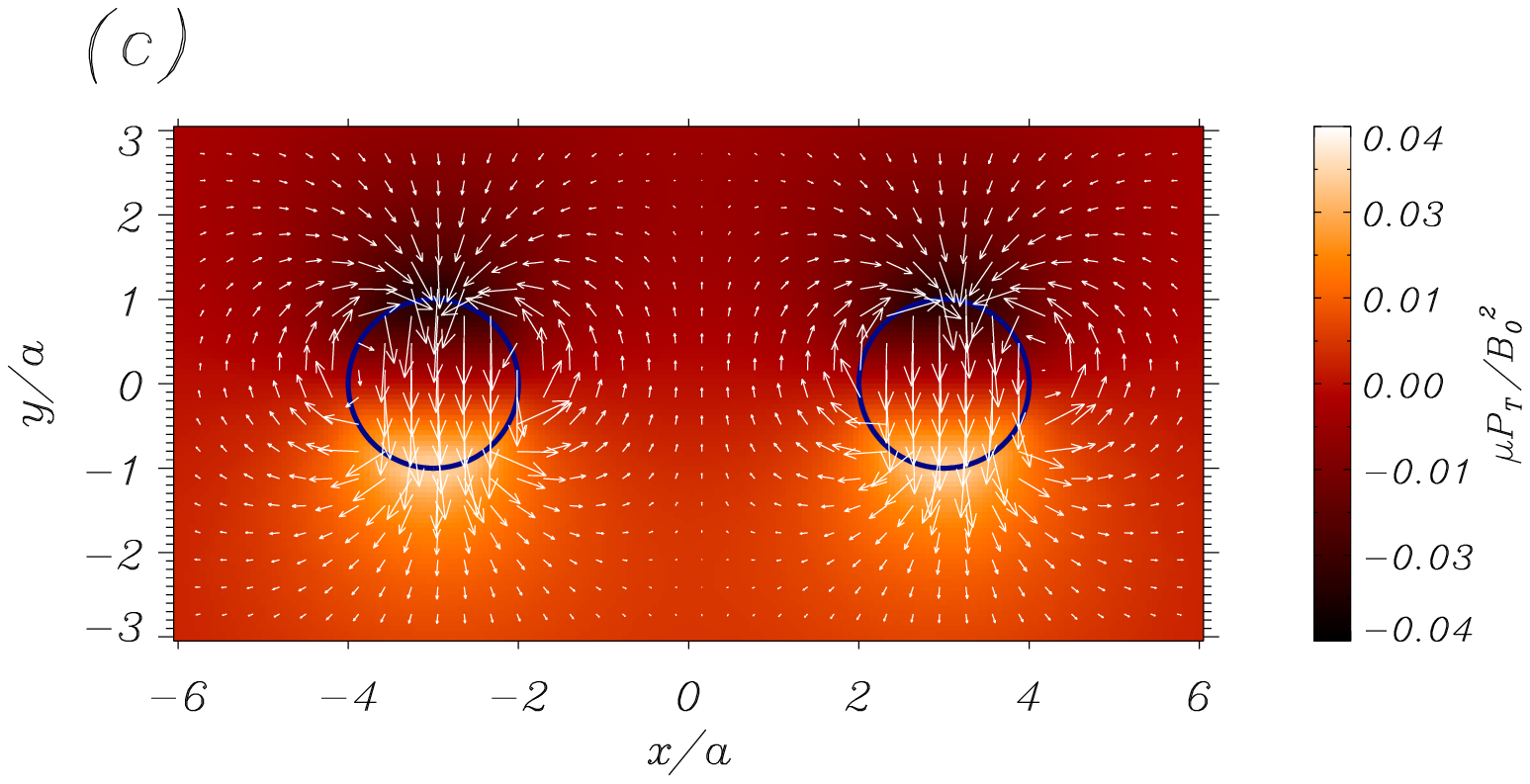}\hspace{0.8cm}\includegraphics[width=7.5cm]{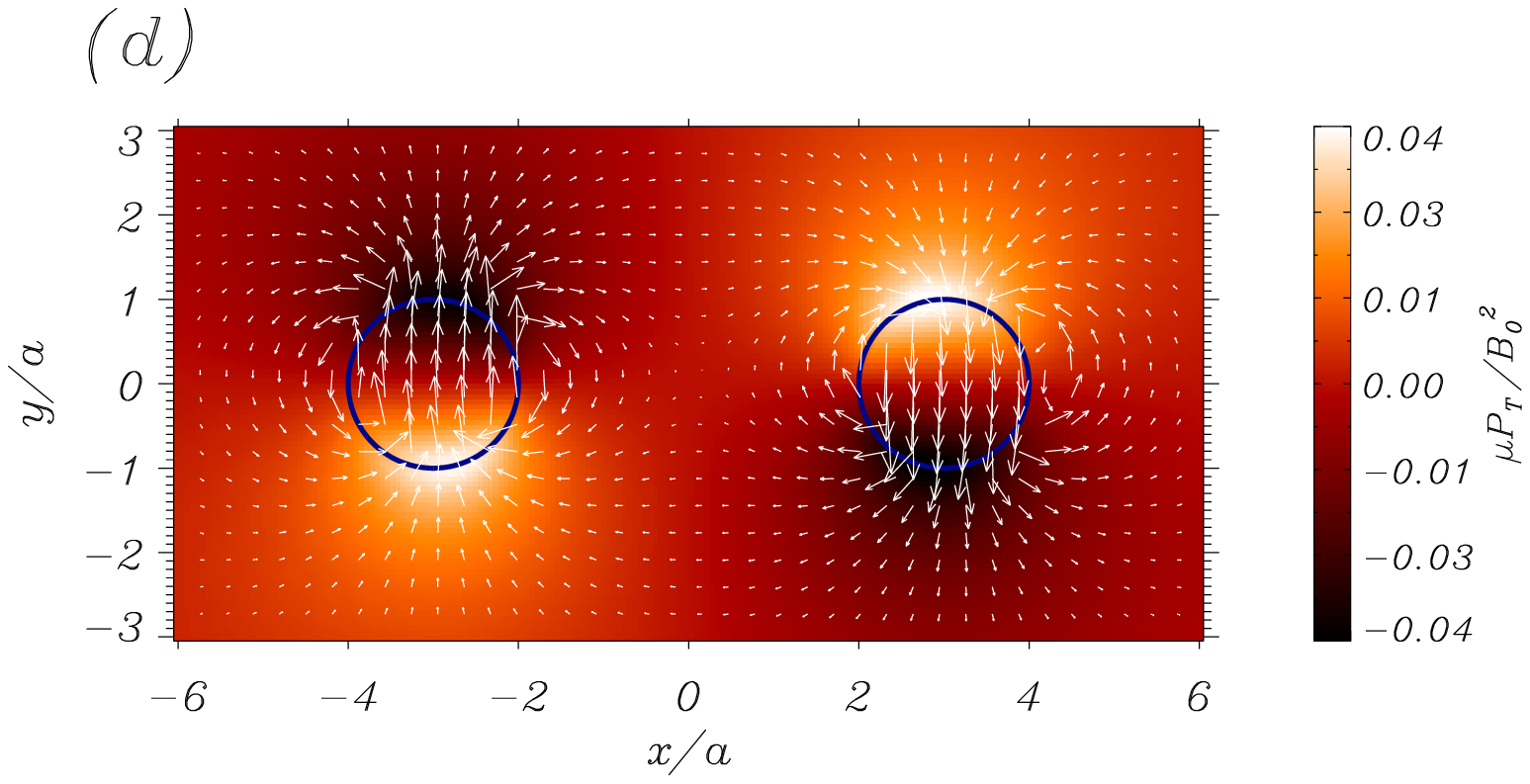}\vspace{-1.cm}
\caption{ 
Total pressure perturbation (color field) and velocity field (arrows) of the
fast four collective normal modes (plotted in the $xy$-plane, see
Fig.~\ref{sketch}).  The modes are labeled as (a) $S_x$, with the loops moving
in phase in the $x$-direction; (b) $A_x$, the tubes move in the $x$-direction
but in antiphase; (c) $S_y$, the tubes move in the $y$-direction in phase; and
finally (d) $A_y$, the loops move in antiphase in the $y$-direction.  Here, the
loop radii are $a=0.1L$ and the distance between centers is $d=6 a$.
}
\label{normal_modes}
\end{figure}

\begin{figure}[!ht]
\center{
{\resizebox{15cm}{!}{\includegraphics{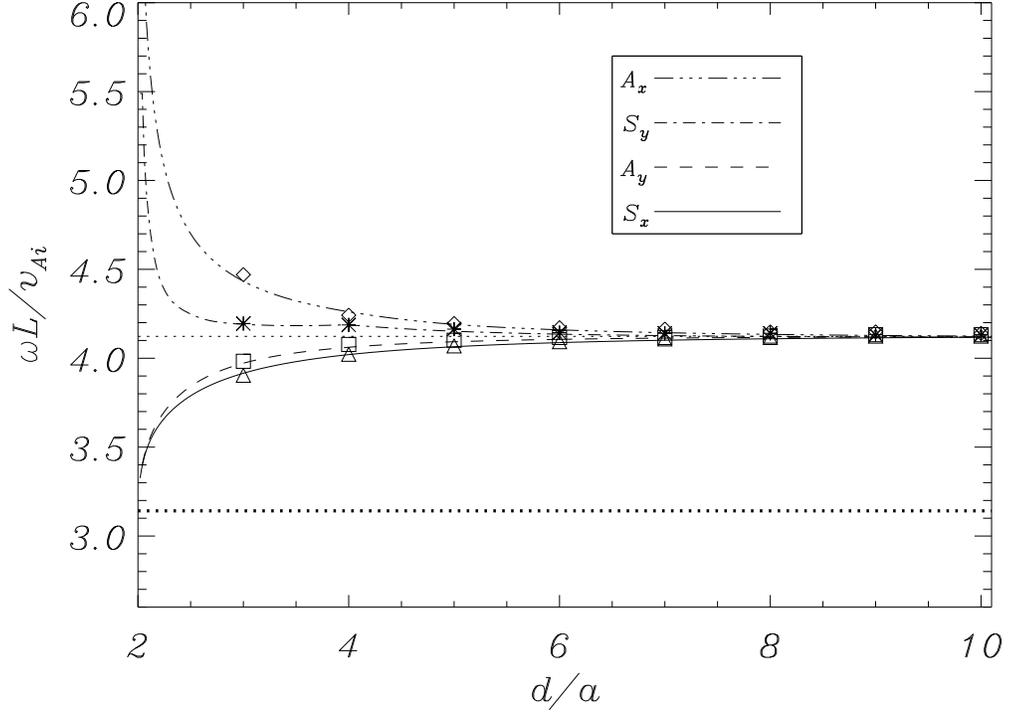}}}
}
\caption{ 
Frequency, $\omega$, as a function of the separation between cylinders, $d$, for
a density enhancement $\rho_\mathrm{i}/\rho_\mathrm{e}=10$ and loop radius
$a=0.1 L$.  The lower horizontal thick dotted line is the internal cut-off
frequency, $\omega_\mathrm{c i} = 3.14/\tau_\mathrm{Ai}$. The thin dotted line
is the kink frequency ($\omega= 4.12/\tau_\mathrm{A i}$) of an individual loop.
The calculated frequencies from the time-dependent results in \S 
\ref{time_dependent_normal_modes} are also plotted as triangles ($S_x$), squares
($A_y$), asterisks ($S_y$), and diamonds ($A_x$).
\label{w_vs_d}
}
\end{figure}

We find four collective fundamental trapped modes (see Fig.~\ref{normal_modes}).
There are other harmonics but we concentrate on the fundamental kink-like modes
because they produce the largest transverse displacement of the loops axes. The
velocity field is more or less uniform in the interior of the loops, and so they
move basically as a solid body, while the external velocity field has a more
complex structure. The four velocity field solutions have a well defined
symmetry with respect to the $y$-axis. In Figure \ref{normal_modes}a, we see
that the velocity field inside the tubes lies in the $x$-direction and is
symmetric with respect to the $y$-axis. We call this mode $S_x$, where $S$
refers to the symmetry of the velocity field and the subscript $x$ refers to the
direction of the velocity inside the tube. The same nomenclature is used for the
other modes. In Figure \ref{normal_modes}b the velocity inside the cylinders is
mainly in the $x$-direction and  antisymmetric with respect to the $y$-axis, so
we call this mode $A_x$. Similarly, in Figure \ref{normal_modes}c the velocity
lies in the $y$-direction and is symmetric with respect to the $y$-axis, while
it is antisymmetric in Figure \ref{normal_modes}d. Hence, we call these modes
$S_y$ and $A_y$, respectively. The pressure field of the $A_x$ and $S_y$ modes
is symmetric  with respect to the $y$-axis, while that of the $S_x$ and $A_y$
modes is antisymmetric.
\clearpage
\clearpage
The frequencies of oscillation of these four modes as a function of the loop
separation, $d$,  are displayed in Figure \ref{w_vs_d}. For large separations
between the tubes, the modes tend to the kink mode of an individual loop (see
dotted line). On the other hand, for smaller separations, they split in four
branches associated to the four oscillatory modes described before. The
splitting effect was noticed in \citet{diaz2005} and \citet{luna} in a
configuration of several slabs. The frequency difference between the modes
increases when the interaction between the loops becomes stronger, i.e.  when
the distance between them is small. When the loops are very close ($d\sim 2a$),
the frequencies of the $S_x$ and $A_y$ modes  tend to the value $\omega=3.33/
\tau_\mathrm{A i}$, which is similar to the internal cut-off frequency,
$\omega_\mathrm{c i}=k_z v_\mathrm{A i}=3.14/ \tau_\mathrm{A i}$ (the difference
is only around 6$\%$). Here $\tau_\mathrm{A i}$ is the Alfv\'en transit time,
defined as $\tau_\mathrm{A i}=L/v_\mathrm{A i}$. On the other hand, in this
limit, the $S_y$ and $A_x$ frequencies are quite large in comparison to the kink
mode frequency.

It is interesting to note that when both tubes move symmetrically in the
$x$-direction, i.e. in the $S_x$ mode, the fluid between follows the loops
motion (see Fig.~\ref{normal_modes}a). On the other hand, when the loops
oscillate antisymmetrically, i.e. in the $A_x$ mode, the intermediate fluid is
compressed and rarefied (see Fig.~\ref{normal_modes}b), producing a more forced
motion than that of the symmetric mode. This is the reason for the $S_x$ ($A_x$)
mode having a smaller (larger) frequency than that of the individual loop. For
the modes polarized in the $y$-direction the behavior is somehow similar,
although in this case the antisymmetric mode (see Fig.~\ref{normal_modes}d) has
a lower frequency than the symmetric mode (see  Fig.~\ref{normal_modes}c). When
one of the loops moves upwards the surrounding fluid near the other loop moves
downwards. This helps to push the other loop in this direction and produces the
antisymmetric motion. The situation is different for the $S_y$ mode, for which
the direction of motion of the surrounding fluid is opposite to that of the
other tube. This explains why the frequency of the $A_y$ solution is smaller
than that of the $S_y$ mode.

\clearpage
\clearpage
\section{Time-dependent analysis: numerical simulations}
\label{time_dependent_analysis}

The initial perturbation that we have used when solving numerically the ideal
MHD equations is a planar pulse in the velocity field of the form
\begin{equation}\label{initial_condition}
\mathbf{v}_\mathrm{0} = \mathbf{K} ~ e^{\left[-\mathbf{K}\cdot
(\mathbf{r}-\mathbf{r}_\mathrm{0}) / \Delta \right]^2} ,
\end{equation}
i.e. a Gaussian profile (of width $\Delta=0.2 L$ centered at
$\mathbf{r}_\mathrm{0} = \left( d/2 , 0, 0\right)$) and direction of propagation
along $\mathbf{K}=-\left( \cos\alpha , \sin \alpha, 0\right)$, $\alpha$ being
the angle between the wavevector  and  the $x$-axis. Here $\mathbf{K}$ also
defines the initial polarization of $\mathbf{v}$, which is perpendicular to the
planar pulse. The initial value of the magnetic field perturbation is zero, and
thus the same applies to the total pressure perturbation. In the
simulations a spatial domain of size $30a \times 30a$ is used and the boundaries
are far from the loops. These boundaries are open, which ensures that the
numerical reflections are negligible.

\begin{figure}[!ht]
\center
\mbox{\hspace{-1.cm}\hspace{8.cm}\hspace{8.cm}}\vspace{-0.cm}
\includegraphics[width=7.5cm]{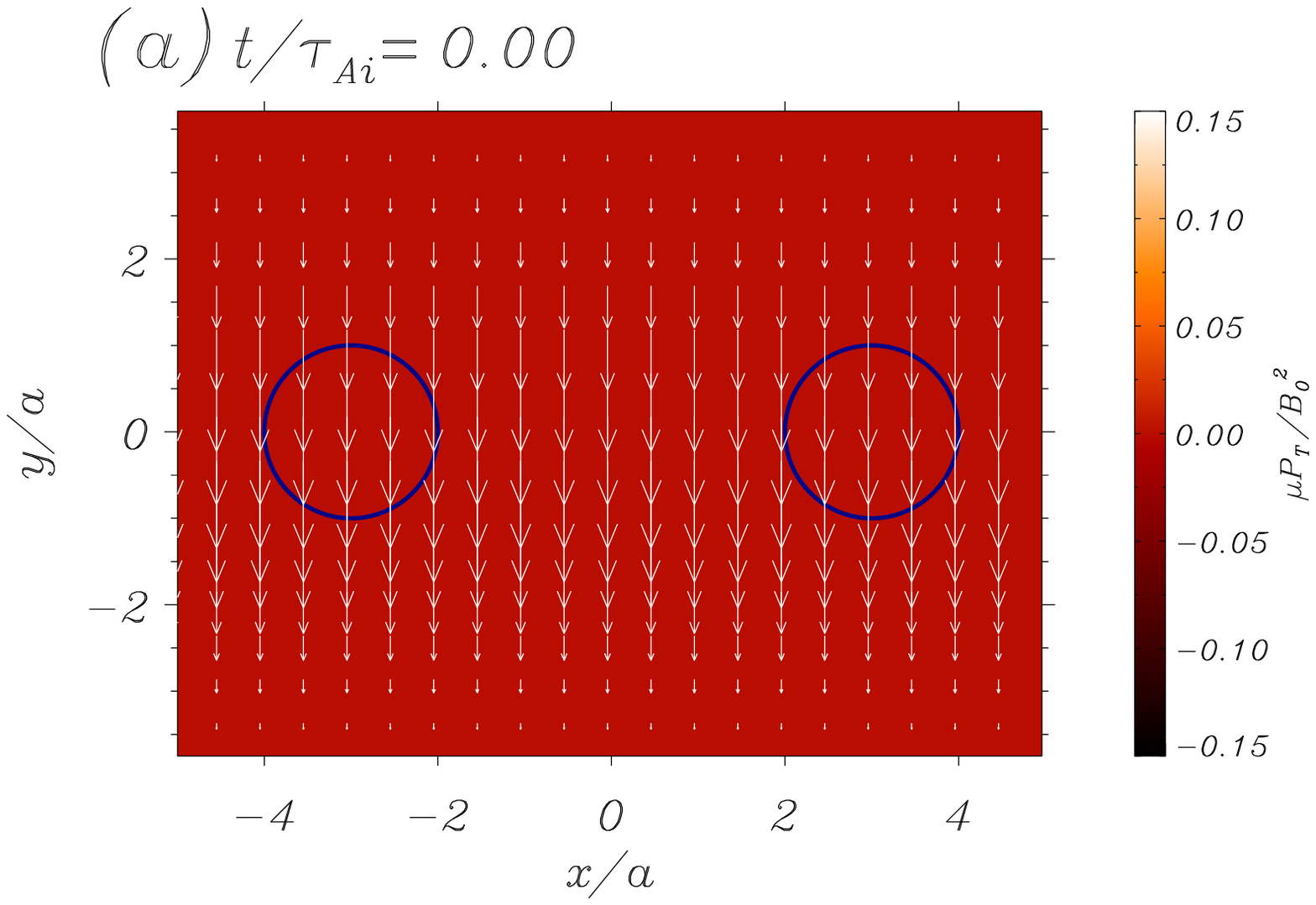}\hspace{0.8cm}\includegraphics[width=7.5cm]{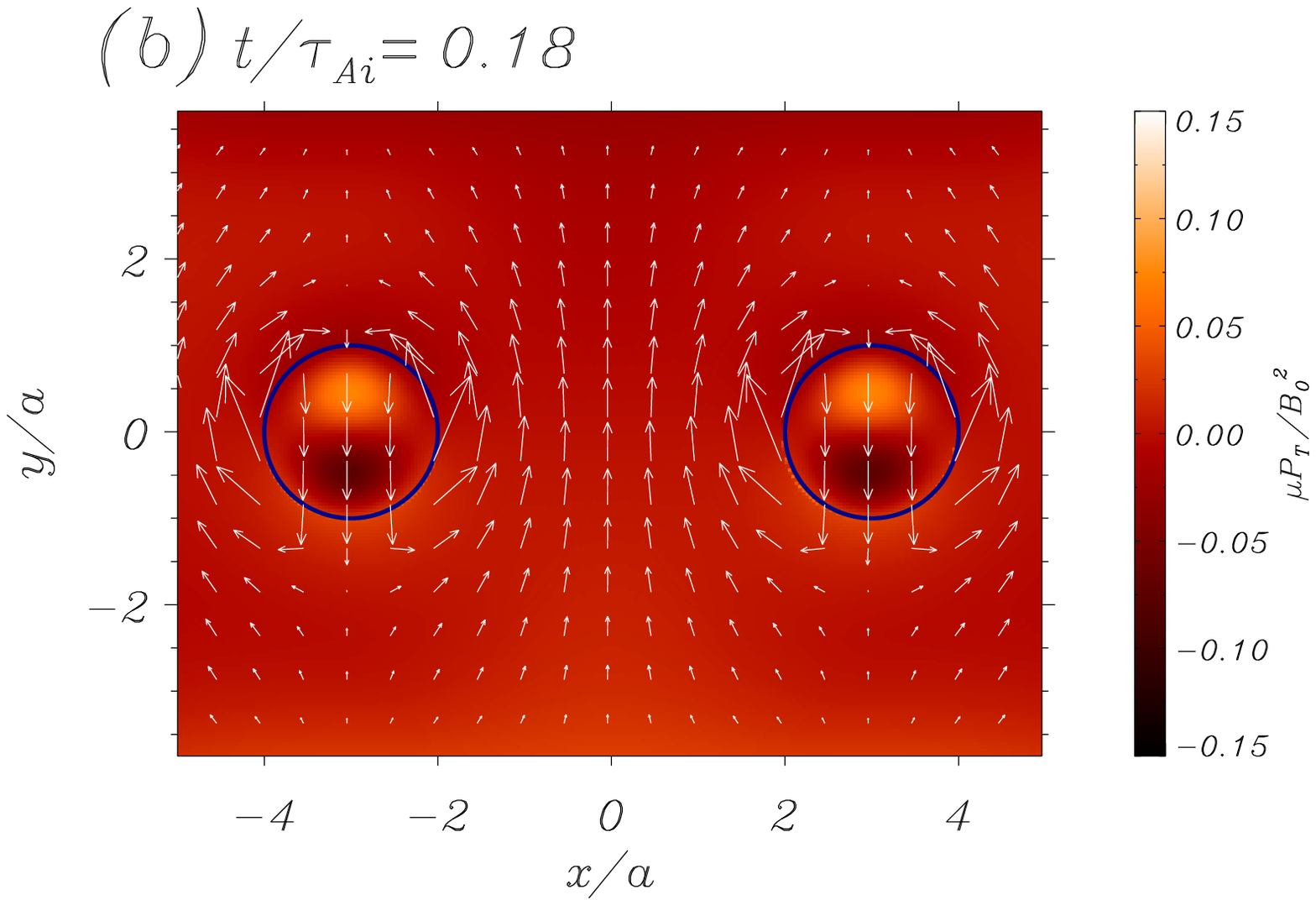}
\mbox{\hspace{-0.4cm}\hspace{8.cm}\hspace{8.cm}}\vspace{-0.5cm}
\includegraphics[width=7.5cm]{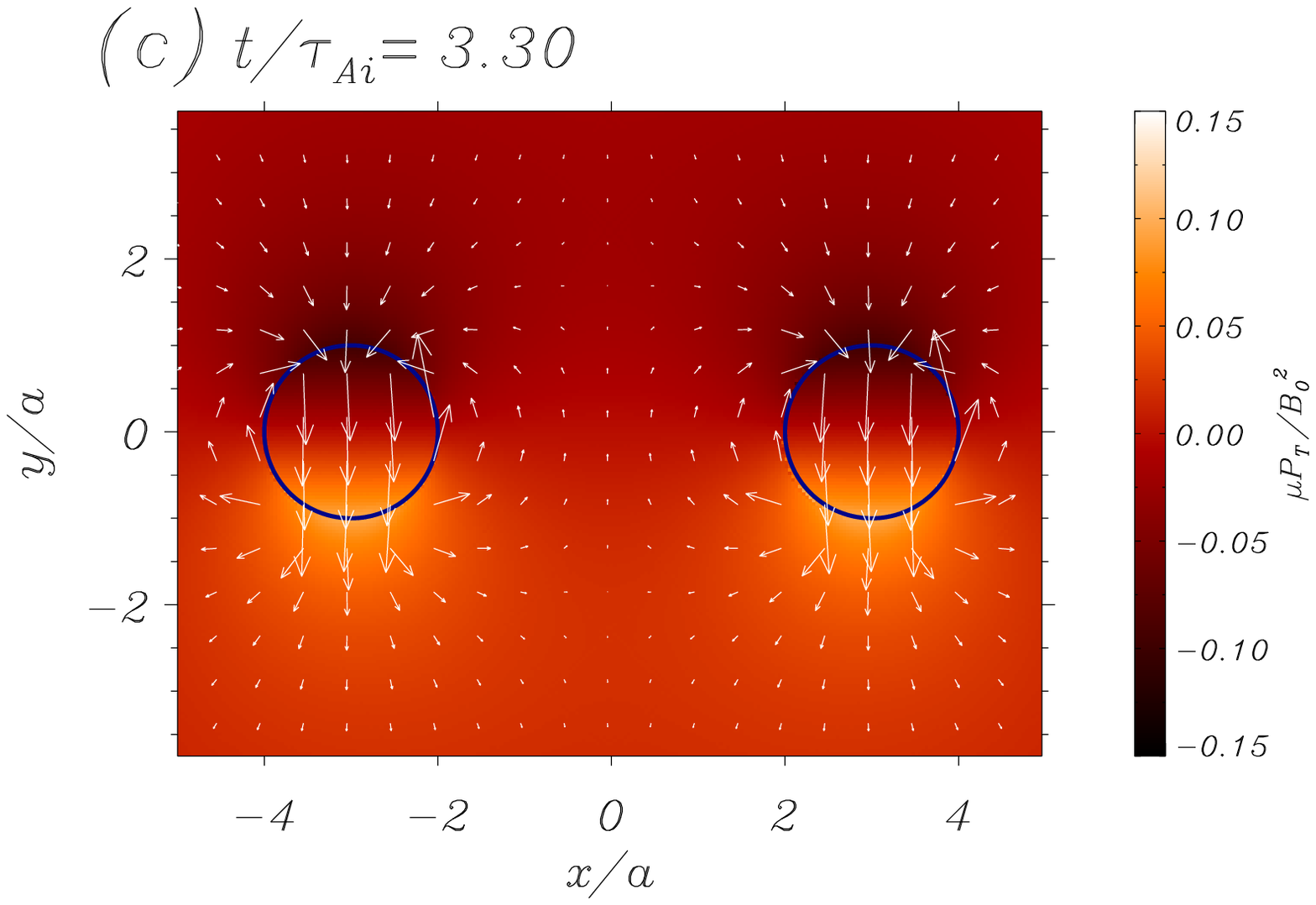}\hspace{0.8cm}\includegraphics[width=7.5cm]{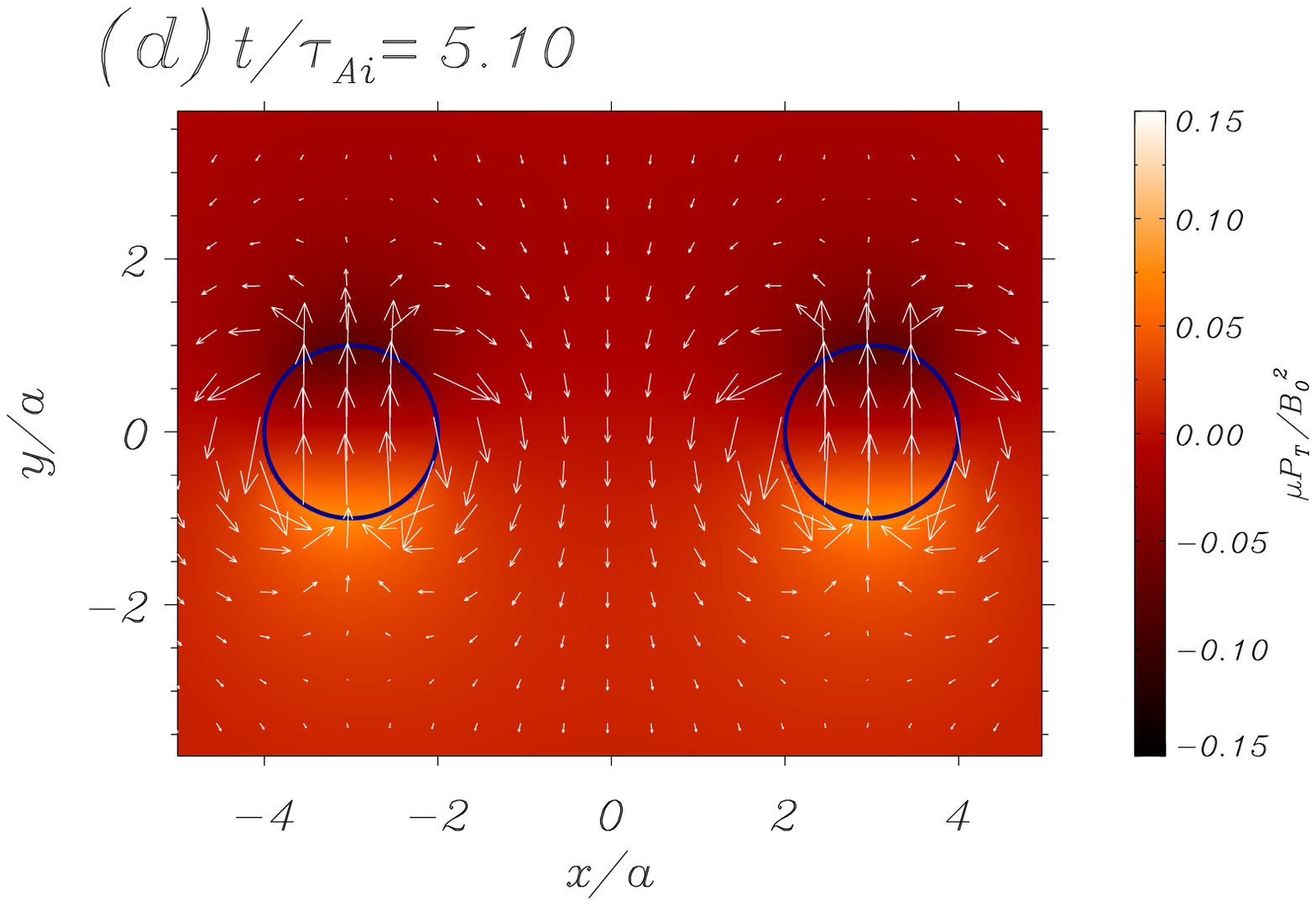}
\caption{
Time-evolution of the velocity field (arrows) and total pressure field (colored
contours), for a separation between loops $d=6 a$ and an initial pulse with an
angle $\alpha=90^\circ$. The two circles mark the positions of the loops at
$t=0$. The panels show different evolution times. In (a) the initial condition
over the velocity field is represented. In (b) the velocity and pressure field
shortly after the initial disturbance, that is, during the transient phase, are
shown. Both tubes are excited at the same time. In panel (b) the tubes are in
the transient phase. In panels (c) and (d) the system oscillates in the
stationary phase with the $S_y$ normal mode. This time evolution is also
available as an mpeg animation in Movie 1. }
\label{t_evol_90}
\end{figure}

\begin{figure}[!ht]
\center
\mbox{\hspace{-1.cm}\hspace{8.cm}\hspace{8.cm}}\vspace{-0.cm}
\includegraphics[width=7.5cm]{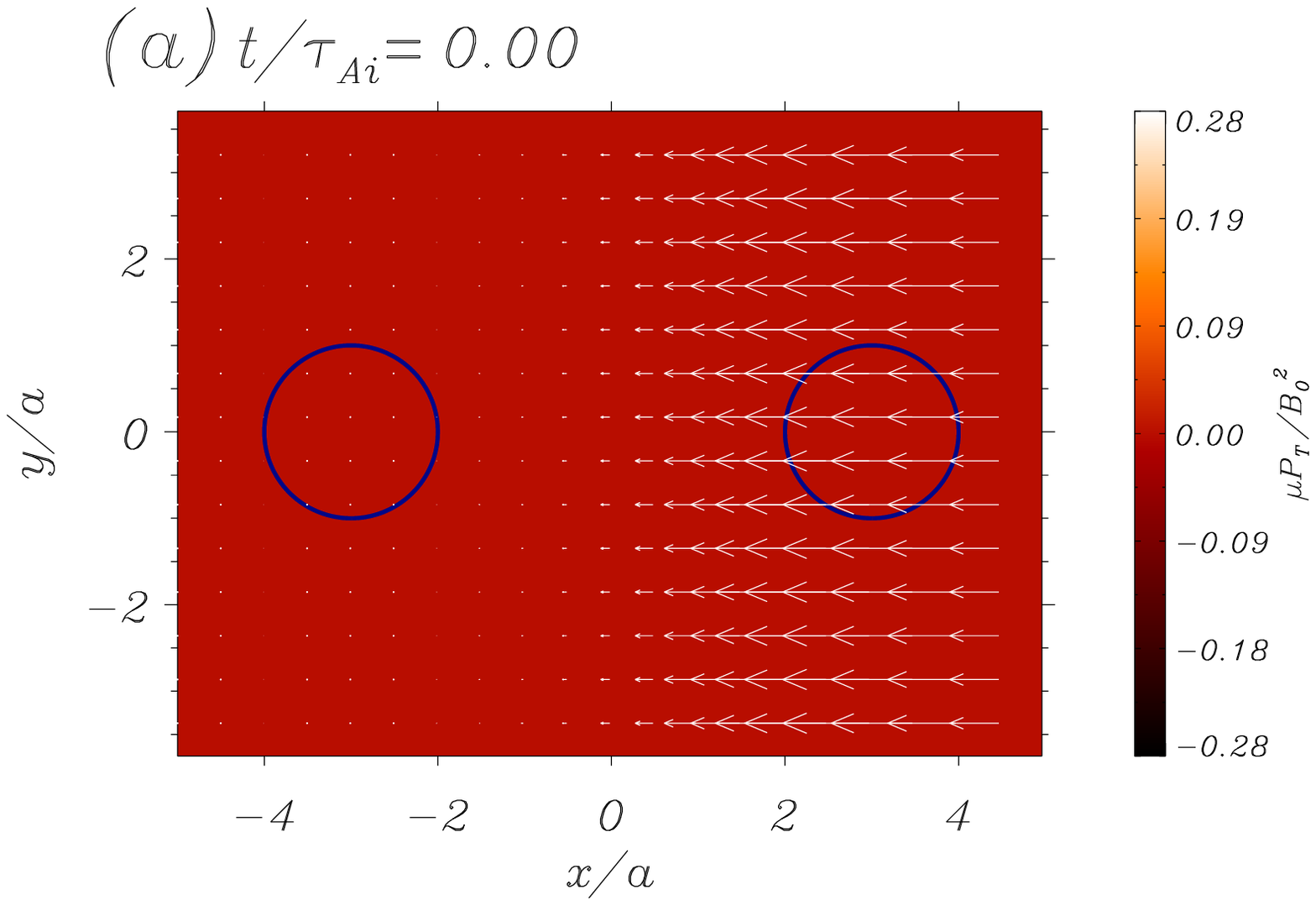}\hspace{0.8cm}\includegraphics[width=7.5cm]{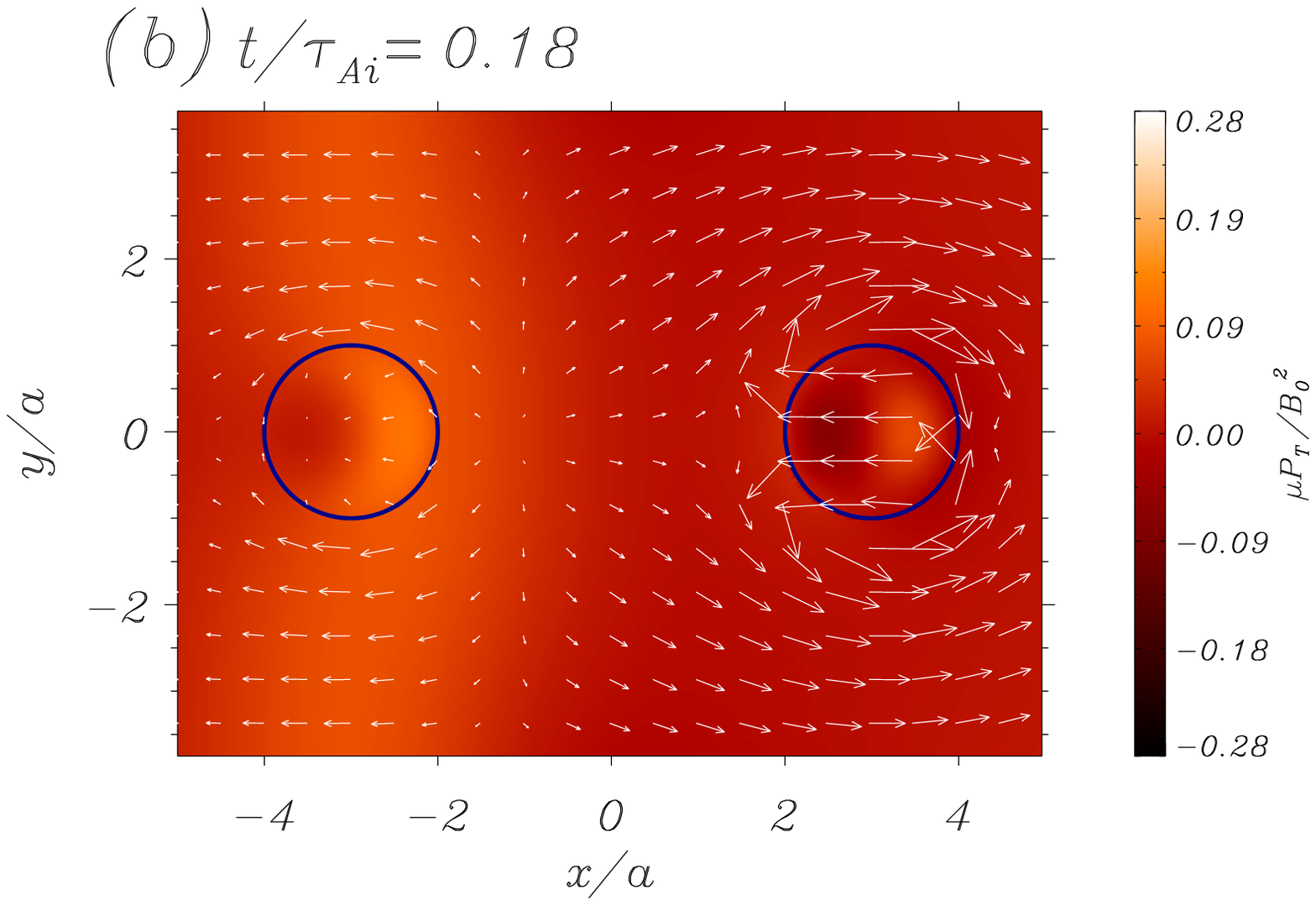}
\mbox{\hspace{-0.4cm}\hspace{8.cm}\hspace{8.cm}}\vspace{-0.5cm}
\includegraphics[width=7.5cm]{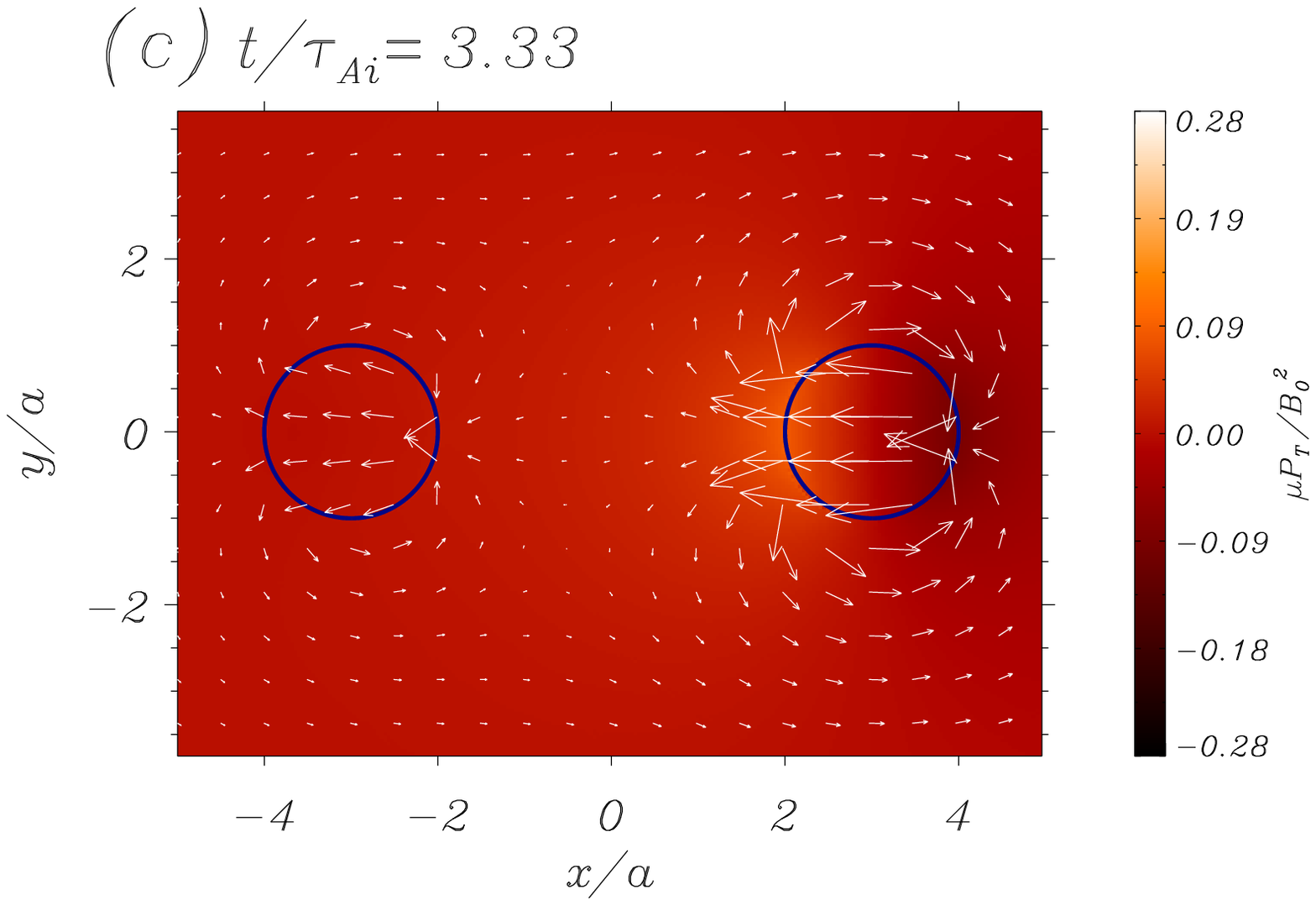}\hspace{0.8cm}\includegraphics[width=7.5cm]{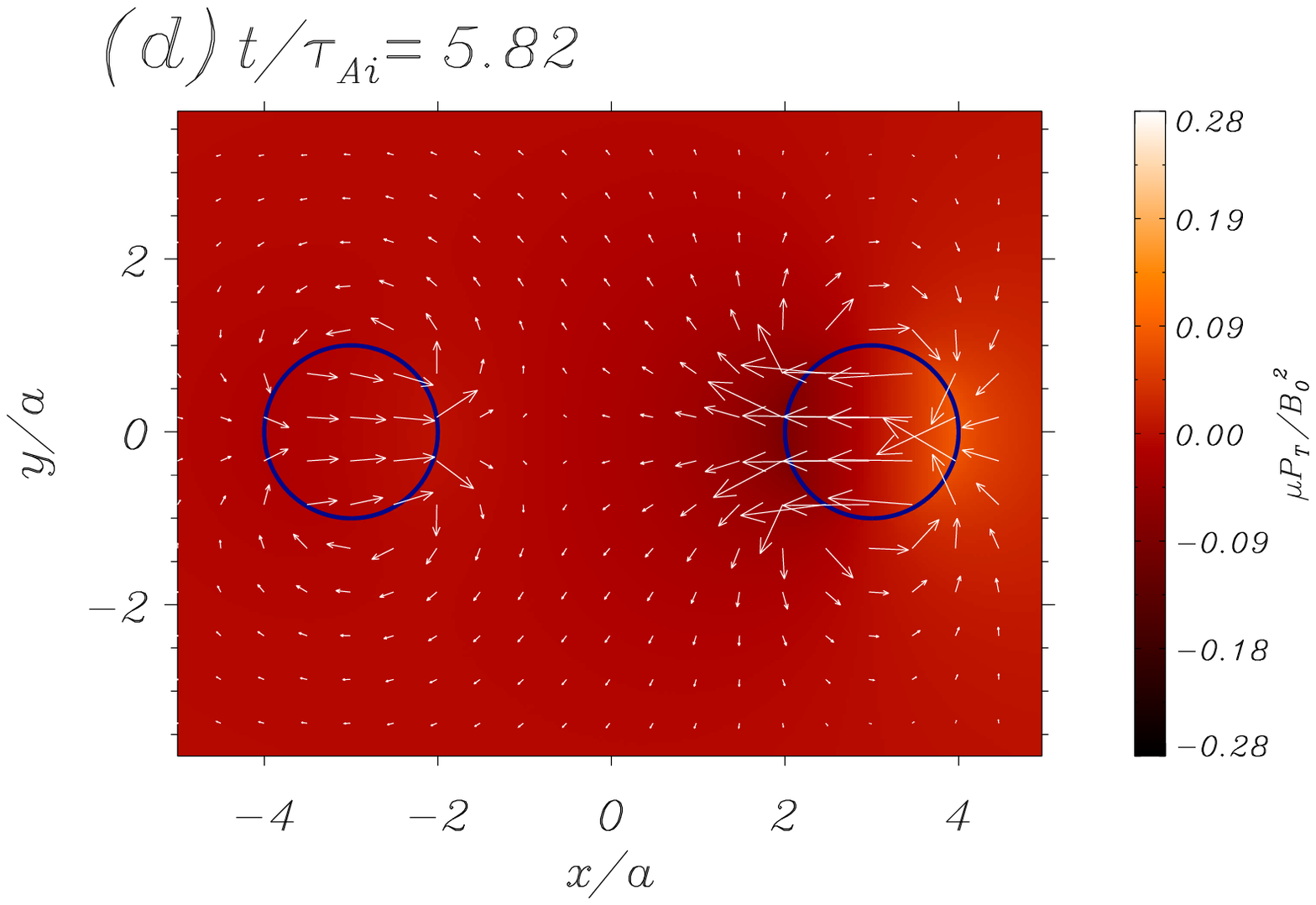}
\caption{
Same as Figure~\ref{t_evol_90} for an initial pulse with an angle
$\alpha=0^\circ$. Here the stationary phase is governed by a superposition of
the $S_x$ and $A_x$ normal modes. The whole time evolution is presented in Movie
2.
}
\label{t_evol_0}
\end{figure}

\begin{figure}[!h]
\center
\mbox{\hspace{-1.cm}\hspace{8.cm}\hspace{8.cm}}\vspace{-0.cm}
\includegraphics[width=7.5cm]{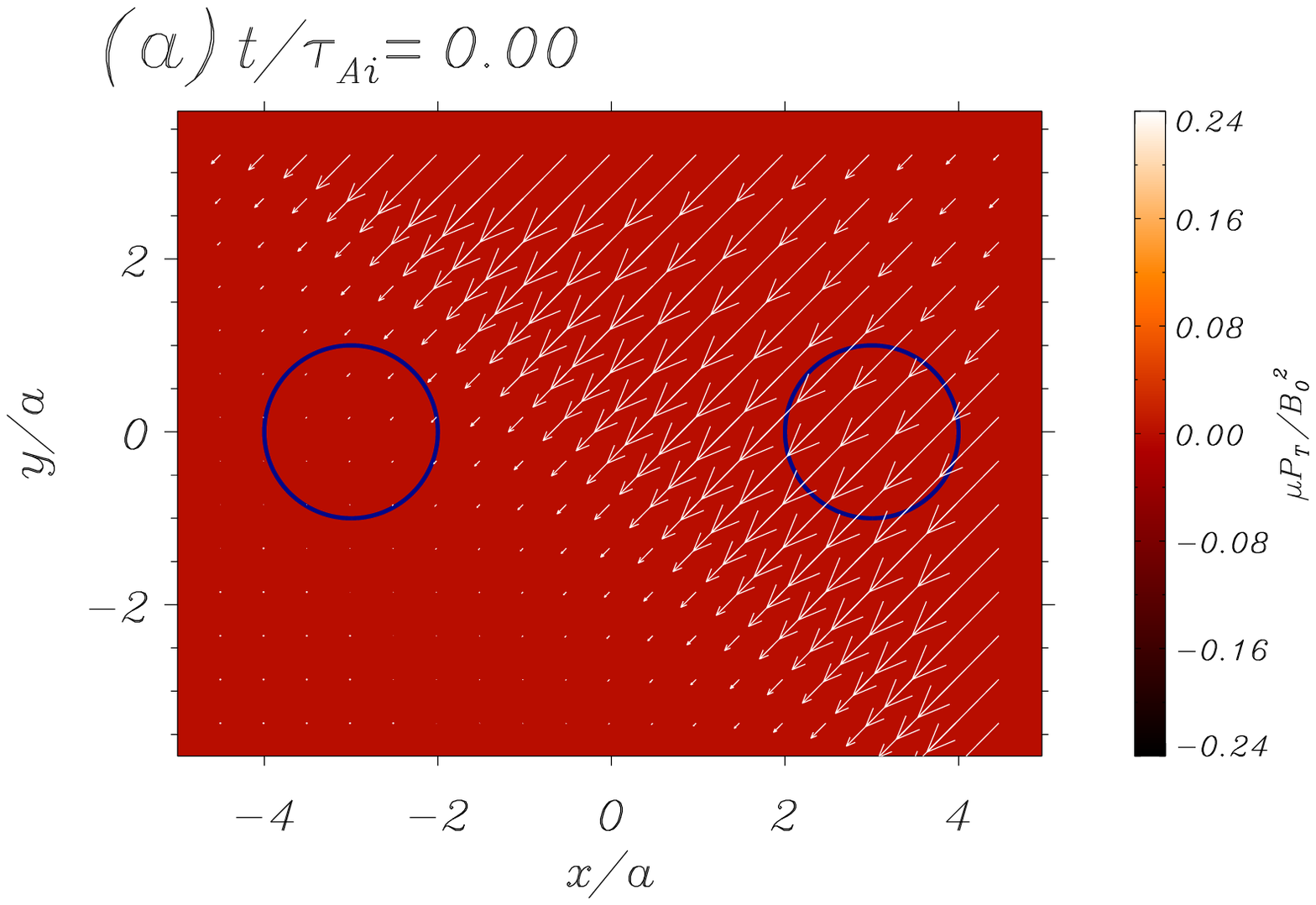}\hspace{0.8cm}\includegraphics[width=7.5cm]{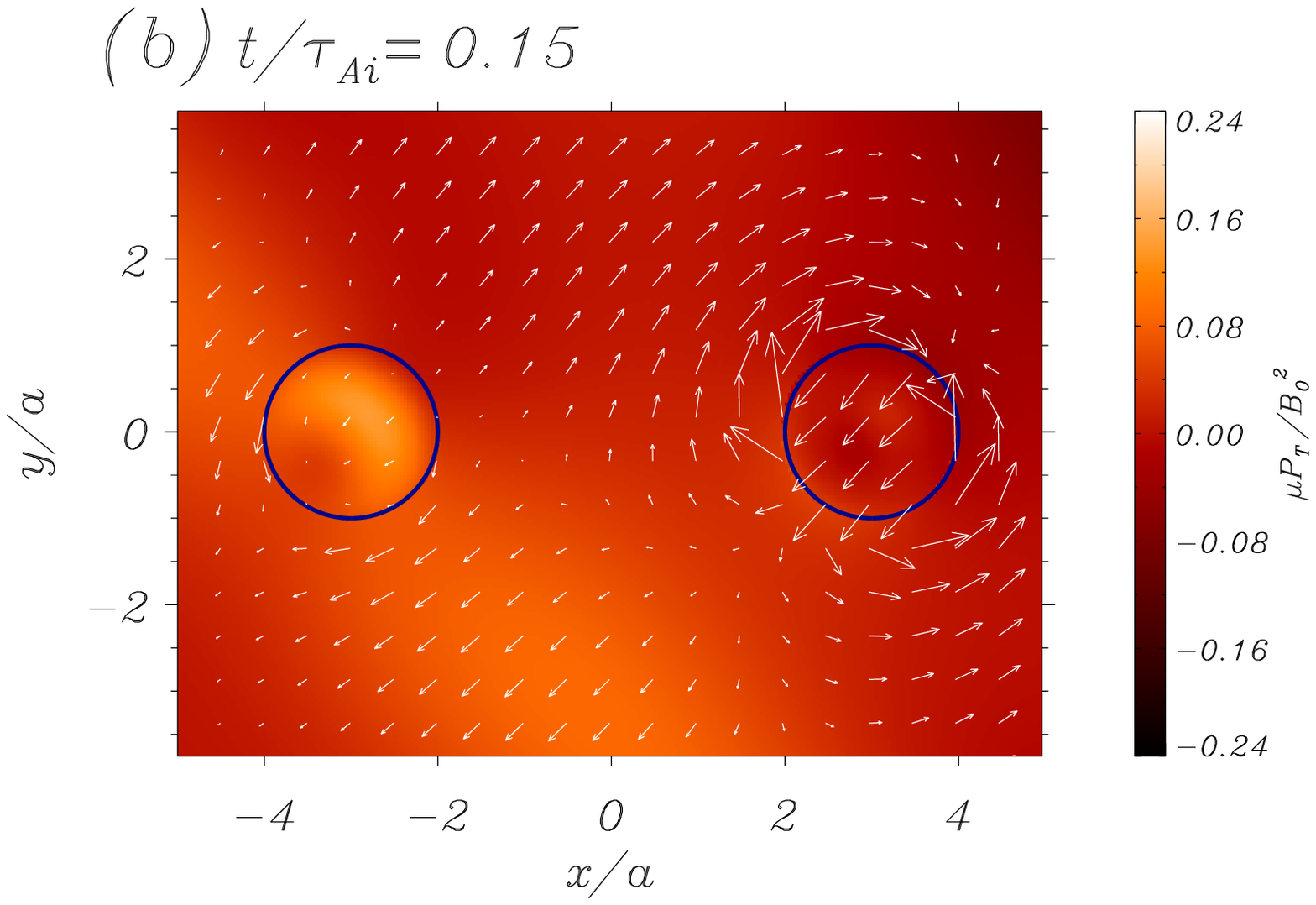}
\mbox{\hspace{-0.4cm}\hspace{8.cm}\hspace{8.cm}}\vspace{-0.5cm}
\includegraphics[width=7.5cm]{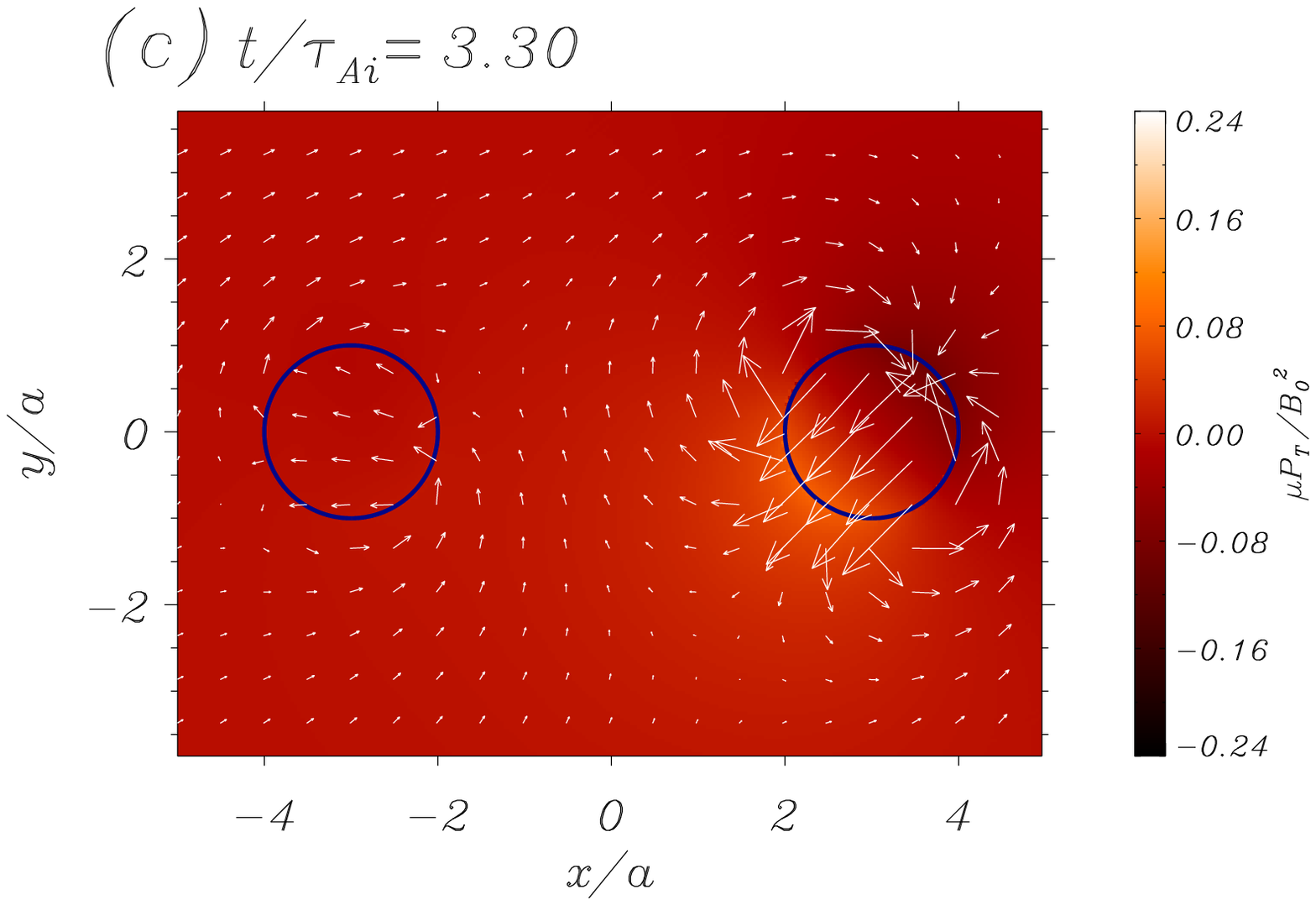}\hspace{0.8cm}\includegraphics[width=7.5cm]{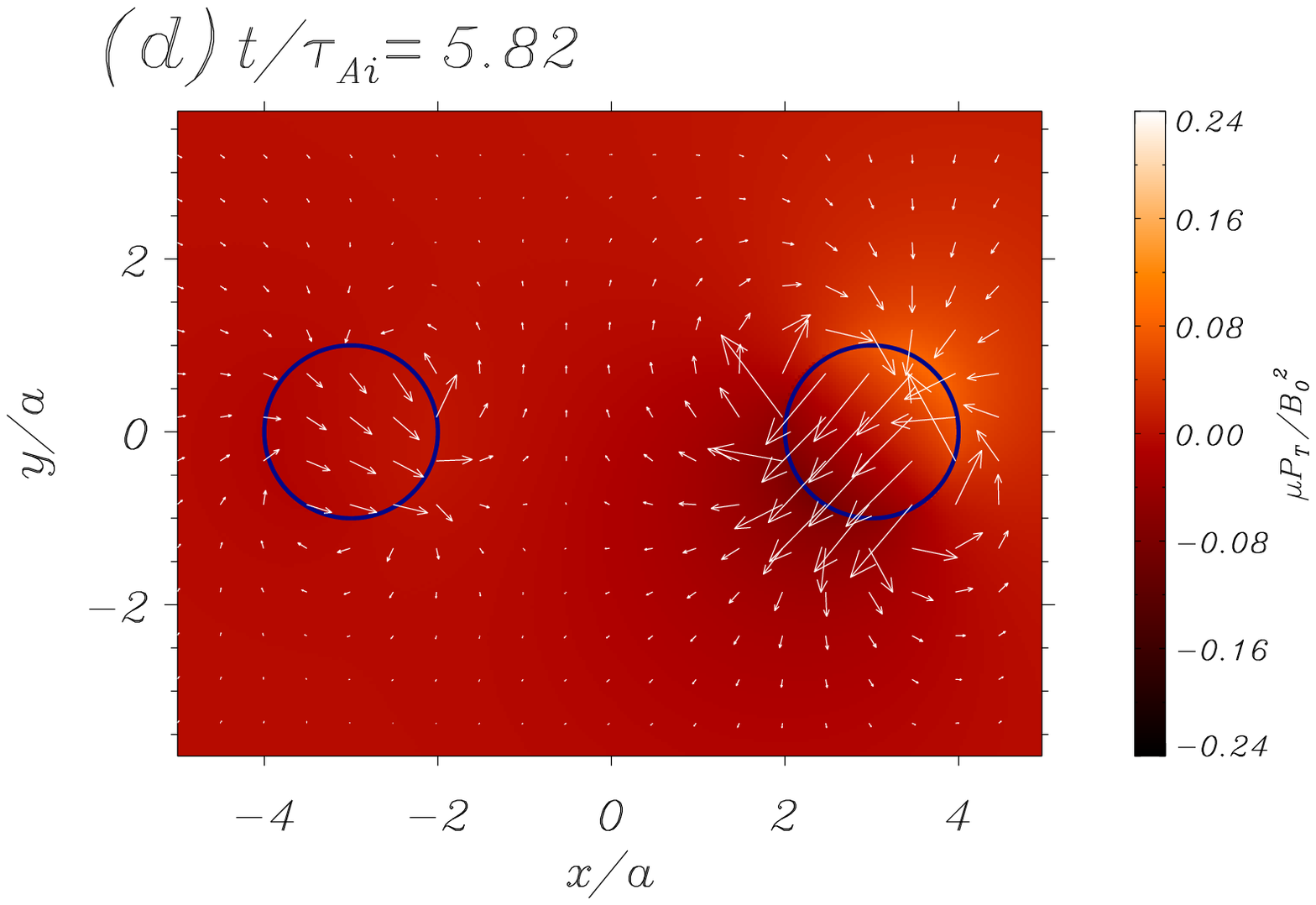}
\caption{
Same as Figure~\ref{t_evol_90} for an initial pulse with an angle
$\alpha=45^\circ$. Here the stationary phase is governed by a superposition of
the $S_x$, $A_x$, $S_y$, and $A_y$ normal modes. The whole time evolution is
presented in Movie 3. Movie 4 contains the time evolution
for much larger times.
}
\label{t_evol_45}
\end{figure}

In Figures \ref{t_evol_90}, \ref{t_evol_0}, and \ref{t_evol_45} three examples
of the time evolution are shown for $\alpha= 90^\circ, 0^\circ$ and $45^\circ$,
respectively, and for a fixed distance between loops $d=6 a$, identical to the
one used in Figure~\ref{normal_modes} (see the time evolution in Movie 1, Movie
2, and Movie 3). 
These three cases illustrate the time evolution of the system after a
perturbation, which consists of two regimes: the transient and the stationary
phases. The stationary phase is characterized by oscillations in one or several
fundamental trapped normal modes (see \S \ref{normal_modes_sec}). On the other
hand, in the transient phase there are leaky modes and internal reflections and
refractions. 
\clearpage
\clearpage
In Figure \ref{t_evol_90} (see Movie 1) the time evolution for the
$\alpha=90^\circ$ initial disturbance is shown, for which, the pulse front lies
along the $x$-axis and excites the $v_y$ component. The loops are perturbed at
the same time (as can be appreciated in Fig.~\ref{t_evol_90}b) and as a
consequence they oscillate symmetrically. In Figure \ref{t_evol_90}b  the system
is in the transient phase, characterized by internal reflections related with
the emission of leaky modes. The external medium has not relaxed yet. Finally,
the system reaches the stationary phase (see Figs. \ref{t_evol_90}c and 
\ref{t_evol_90}d)  and oscillates with the $S_y$ trapped mode (compare the
velocity field and the pressure distribution with Fig.~\ref{normal_modes}c). 

\clearpage
\clearpage

In Figure \ref{t_evol_0} (and Movie 2), the time evolution
for the $\alpha=0^\circ$ initial disturbance is shown. Now the pulse is centered
on the right loop (see Fig.~\ref{t_evol_0}a) and excites the $v_x$ component. In
Figure \ref{t_evol_0}b, the pulse reaches the left tube and passes through it,
the system still being in the transient phase. On the other hand, in Figures
\ref{t_evol_0}c and \ref{t_evol_0}d the system oscillates in the stationary
phase. It is interesting to note that this particular initial disturbance does
not excite the left loop; neither at $t=0$ nor during the transient phase. 
Nevertheless, the oscillatory amplitude in the left loop grows with time in the
stationary phase,  while the amplitude in the right loop decreases in the time
interval shown in Figures \ref{t_evol_0}c and \ref{t_evol_0}d (see also Movie
2). Then, it is clear that the left tube acquires its movement through the
interaction with the right loop, i.e. by a transfer of energy from the right
loop to the left loop. This process is reversed and repeated periodically: once
the left loop has gained most of the energy retained by the loops system, so
that the right loop is almost at rest, the left tube starts giving away its
energy to the right cylinder, and so on. This is simply a beating phenomenon,
that can be explained  in terms of the normal modes excited in this numerical
simulation. In fact, the initial disturbance excites the $S_x$ and $A_x$ modes
with the same amplitude and for this reason  the excitation is initially maximum
on the right tube and zero on the left tube. A more detailed discussion about
this issue is given in \S \ref{2d_beating}.

\clearpage
\clearpage

Finally, we discuss the results for an excitation with $\alpha=45^\circ$. This
simulation is the most complex and general of all (see Movie 3).
As we can see in Figure \ref{t_evol_45}a now both components of the velocity are
excited. In Figure \ref{t_evol_45}b the initial pulse reaches the left tube and
passes through it, but only leaky modes are excited. In Figures \ref{t_evol_45}c
and \ref{t_evol_45}d the system oscillates in the stationary phase, which is a
combination of the four modes $S_x$, $A_x$, $S_y$ and $A_y$. As in the previous
case, there is beating but now it is present in both the $x$- and $y$-velocity
components.  Like for the previous simulation, the left loop is almost still
until the stationary phase (see also dotted curves in
Figs.~\ref{time_evol_example_1}a and \ref{time_evol_example_1}d) despite that in
this simulation the pulse directly hits the left loop without the obstacle of
the right loop. In \S \ref{2d_beating} details about the behavior of the system
are given.

Once we know the general features of the excitation of the two cylinders we can
perform a parametric study of the effect of the distance between the loops and
also the angle of excitation on the loops motion. 

\subsection{Effect of the distance between loops}

\label{time_dependent_normal_modes}

We generate an initial disturbance with an angle of $45^\circ$ for different
distances $d$ and measure the velocity in the loops as a function of time. From
this information we can extract the frequencies of oscillation. As we have seen,
since the velocity field inside the loops is more or less uniform (see
Fig.~\ref{normal_modes}), it is enough to measure the velocity at the center of
the loops to describe their global motion. The reason for choosing the initial
disturbance with $\alpha=45^\circ$ is that it excites the four normal modes, so
that with a single simulation we can measure their frequencies.

\clearpage
\begin{figure}[!ht]
\center
\mbox{\hspace{-0.cm}\hspace{4.cm}\hspace{4.cm}}\vspace{-0.cm}
\includegraphics[width=4.5cm]{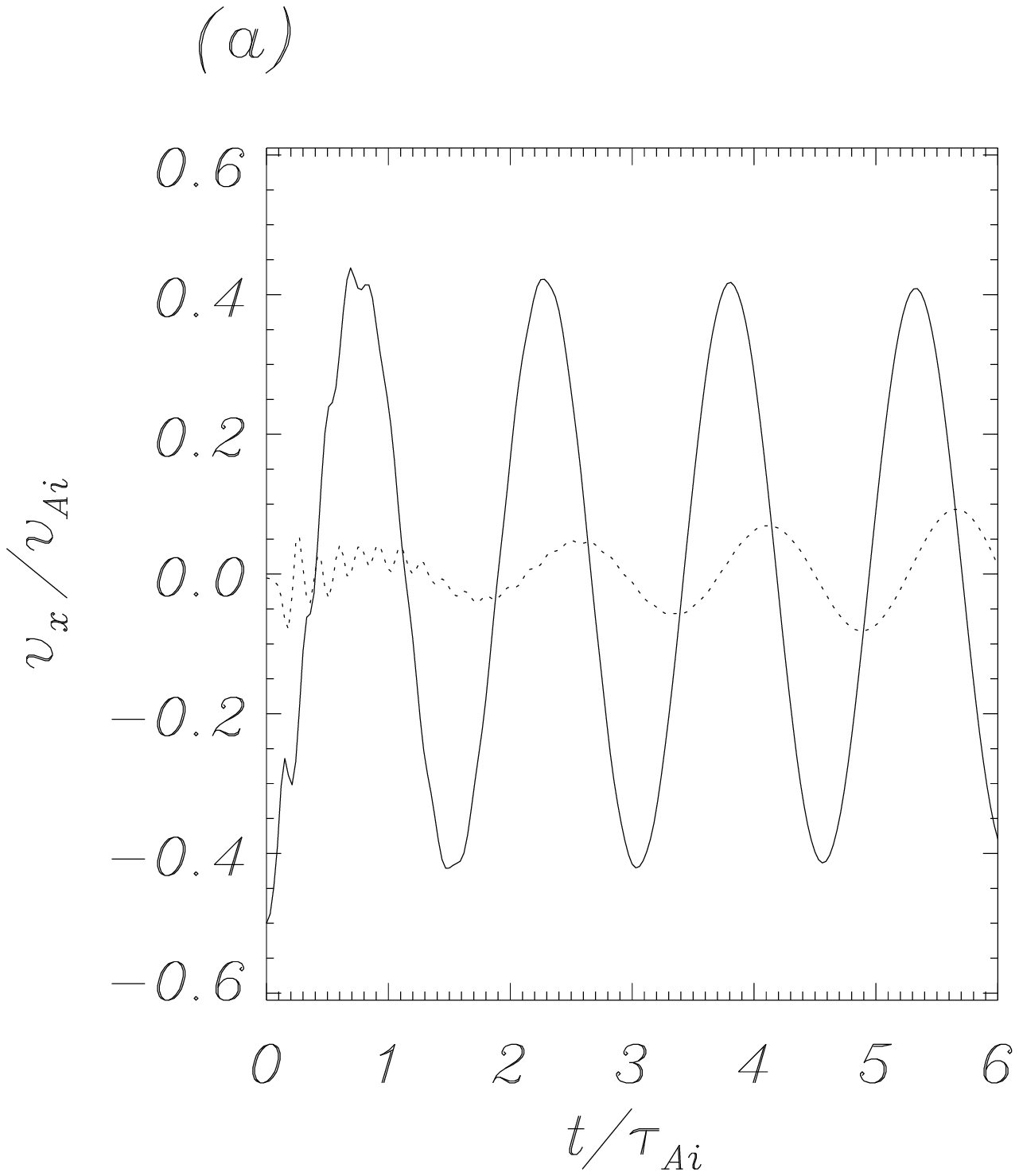}\includegraphics[width=4.5cm]{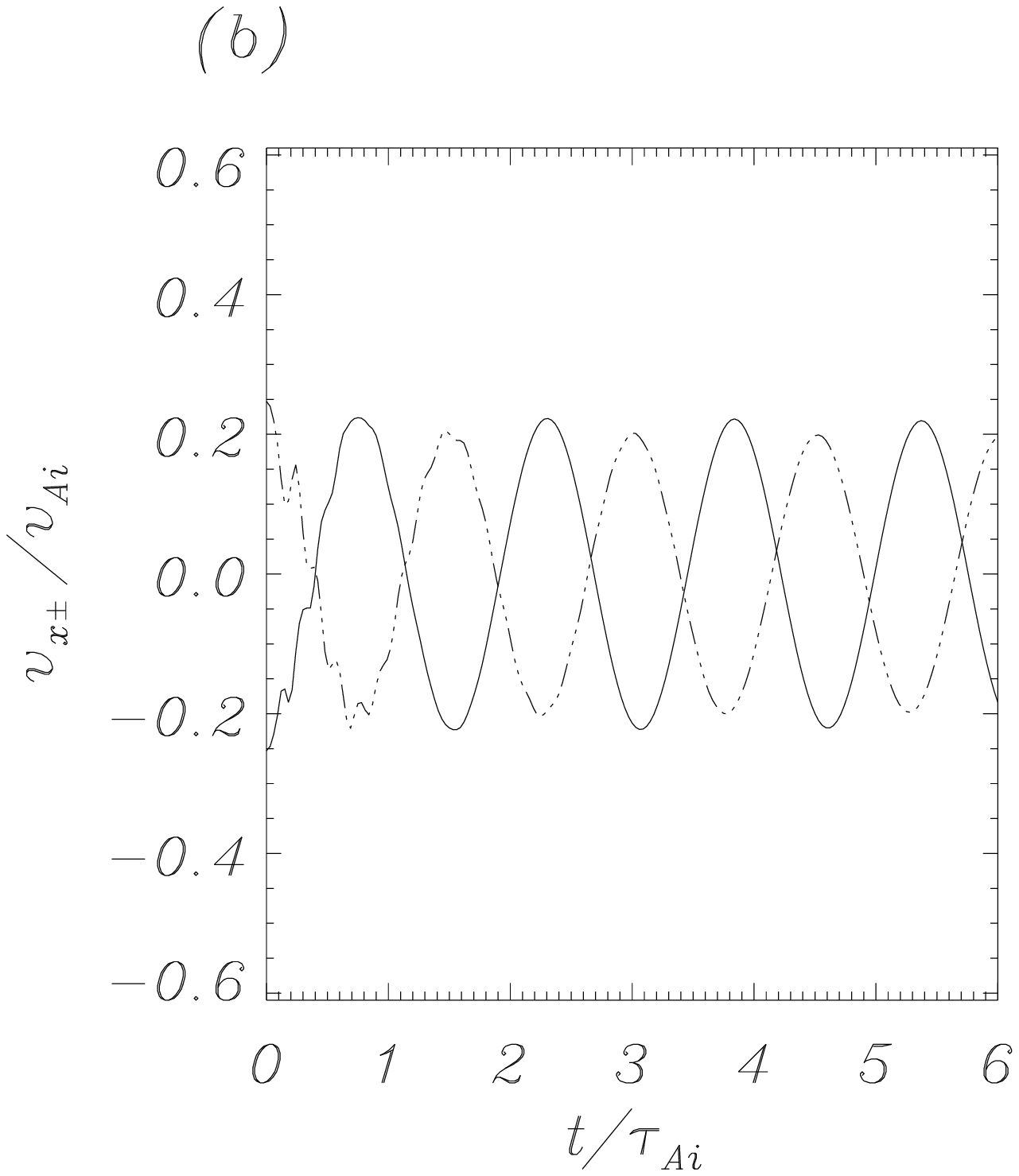}\includegraphics[width=4.5cm]{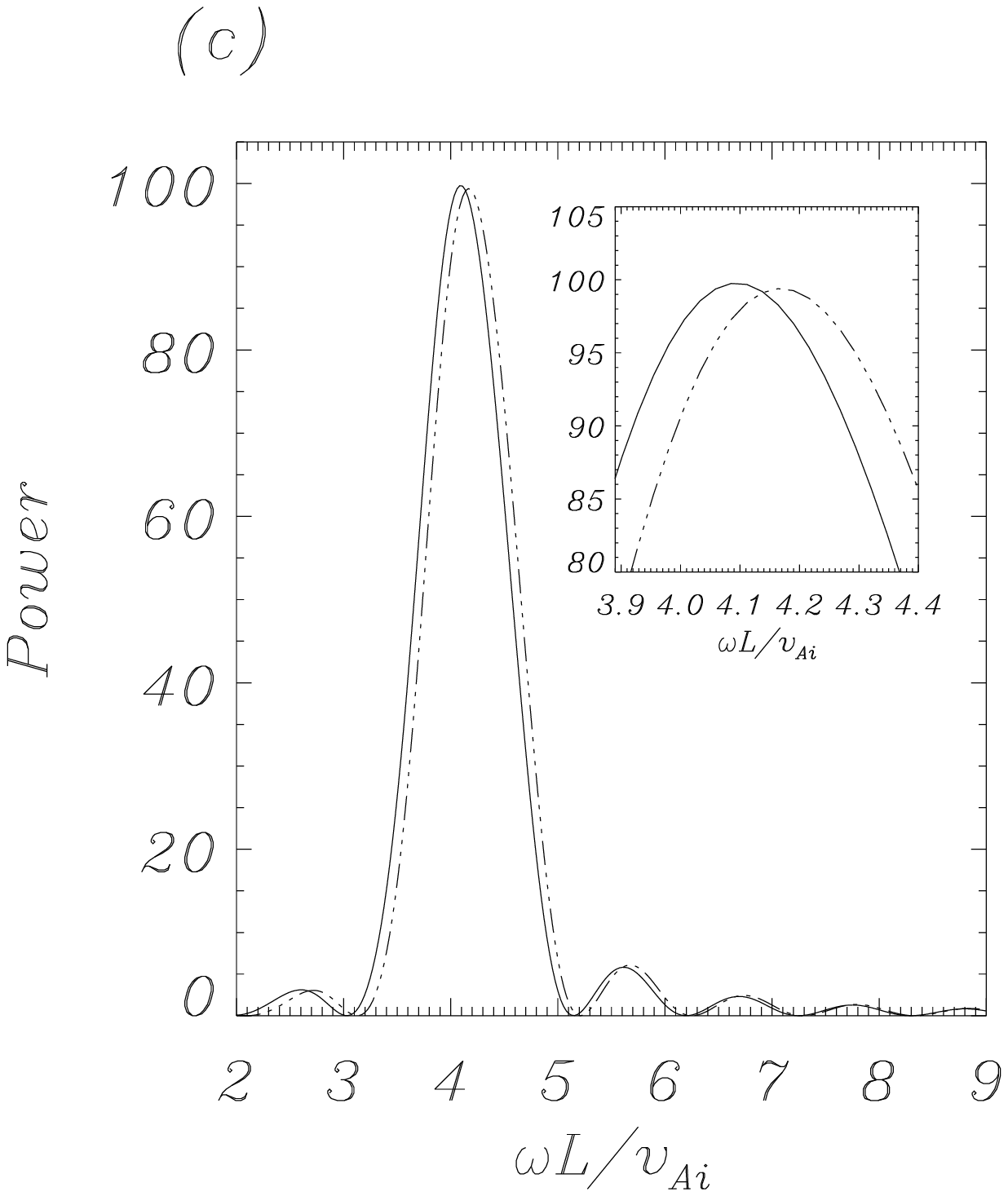}
\mbox{\hspace{-0.cm}\hspace{4.cm}\hspace{4.cm}}\vspace{-0.8cm}
\includegraphics[width=4.5cm]{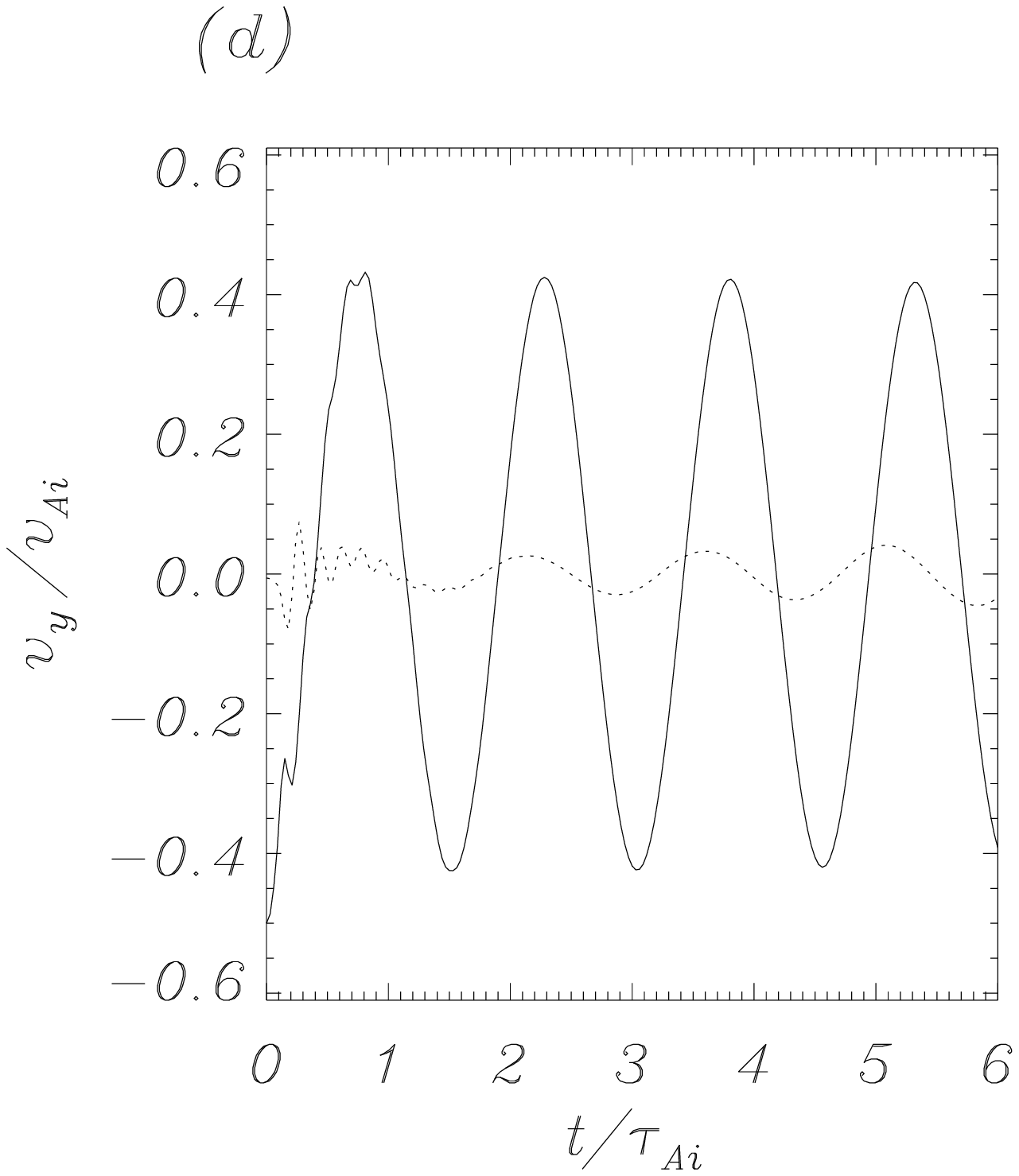}\includegraphics[width=4.5cm]{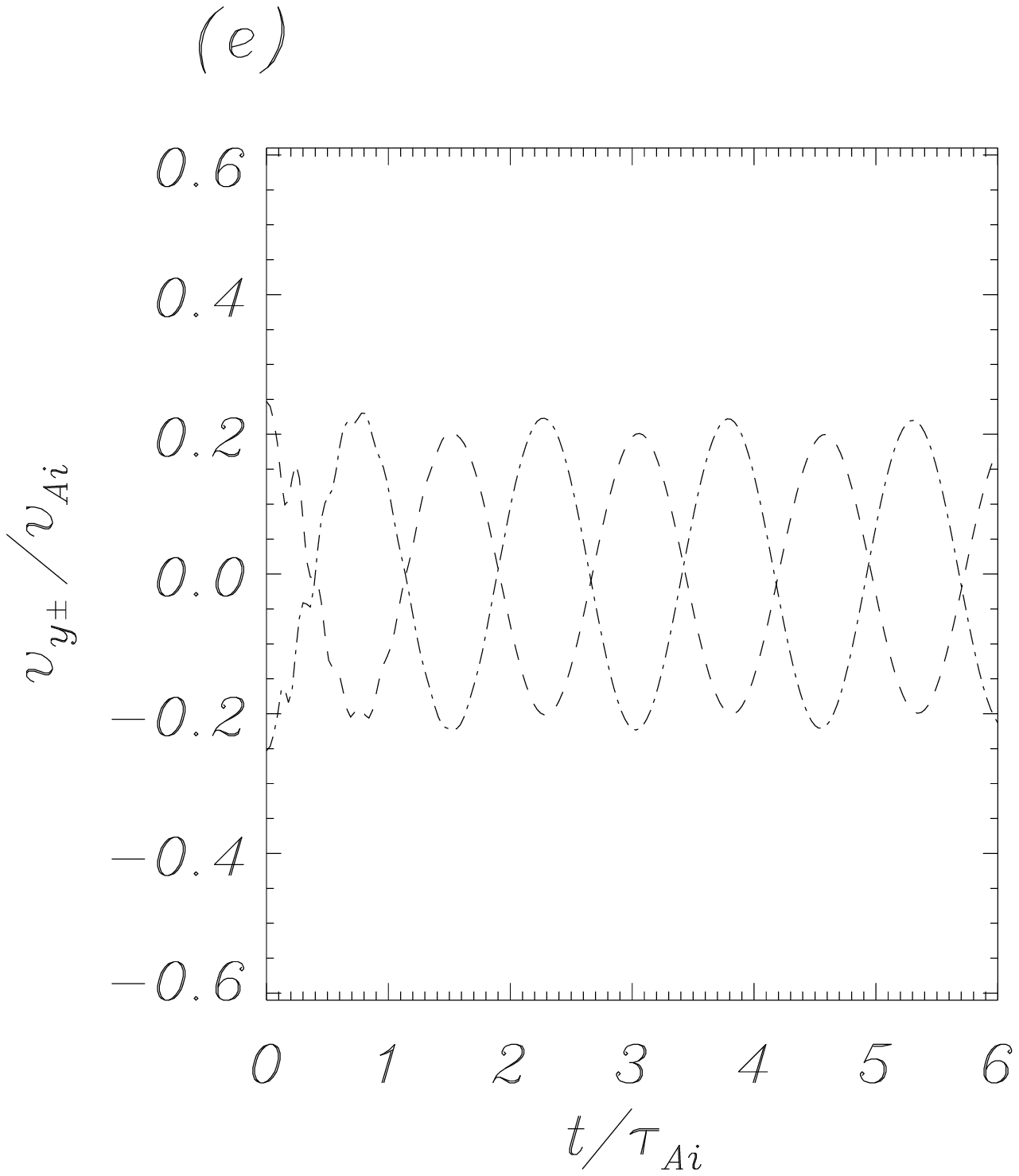}\includegraphics[width=4.5cm]{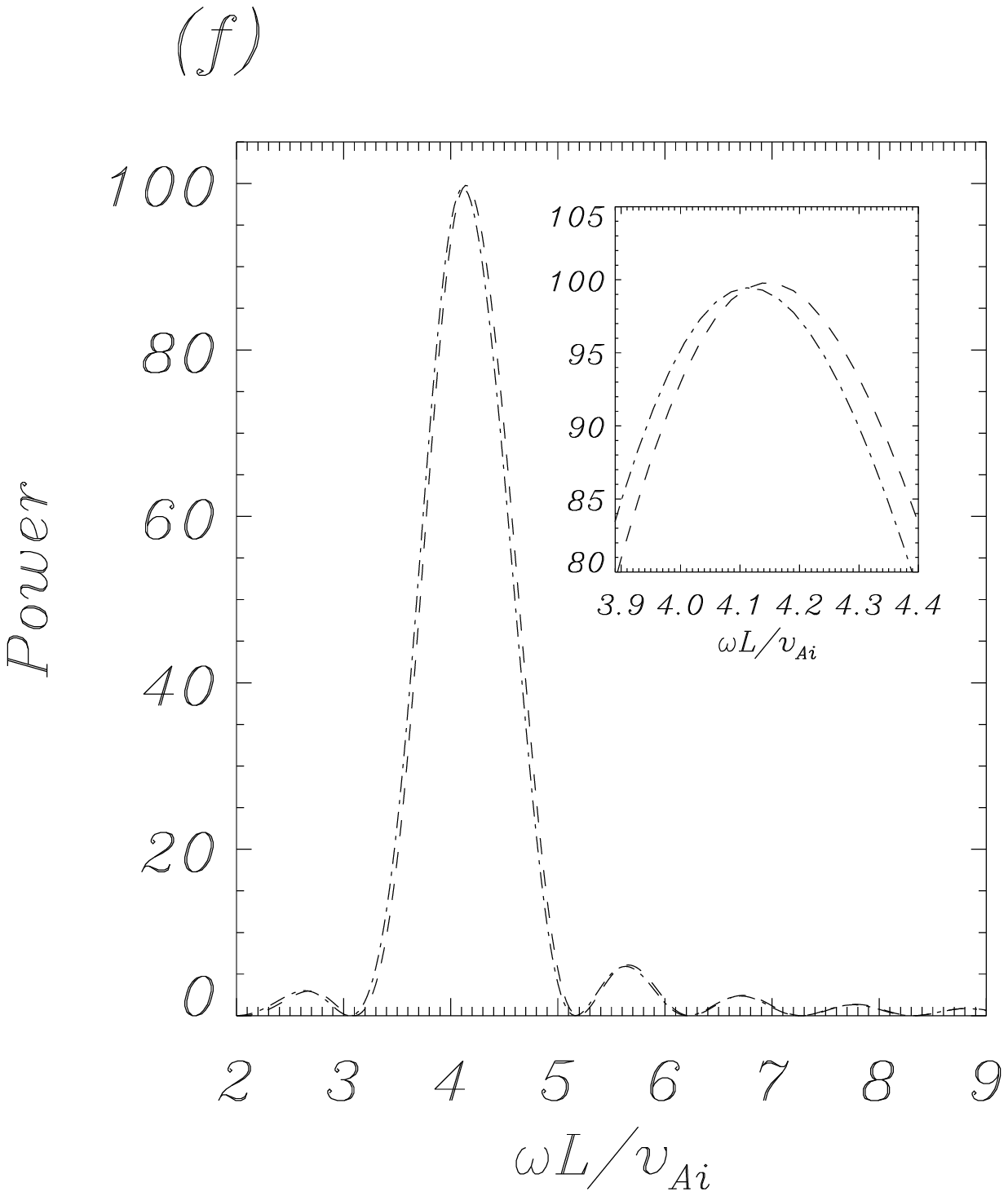}
\caption{
(a) $x$-component and (d) $y$-component of the velocity at the center of the
right (solid line) and left (dotted line) loops for the numerical simulation of
Figure~\ref{t_evol_45} (i.e. with an initial incidence angle $\alpha=45^\circ$).
With the method explained in \S \ref{time_dependent_normal_modes} the normal
mode velocities are extracted and plotted in (b) for the $S_x$ (solid line) and
the $A_x$ (three-dot-dashed line) modes and in (e) for the $S_y$ (dashed line)
and $A_y$ (dot-dashed line) modes. The corresponding power spectra are plotted
with the same line styles in (c) and (f). Power maxima allow us to determine the
frequency of the normal modes from the numerical simulation.
}
\label{time_evol_example_1}
\end{figure}

\begin{figure}[!ht]
\center
\mbox{\hspace{-0.cm}\hspace{4.cm}\hspace{4.cm}}\vspace{-0.1cm}
\includegraphics[width=4.5cm]{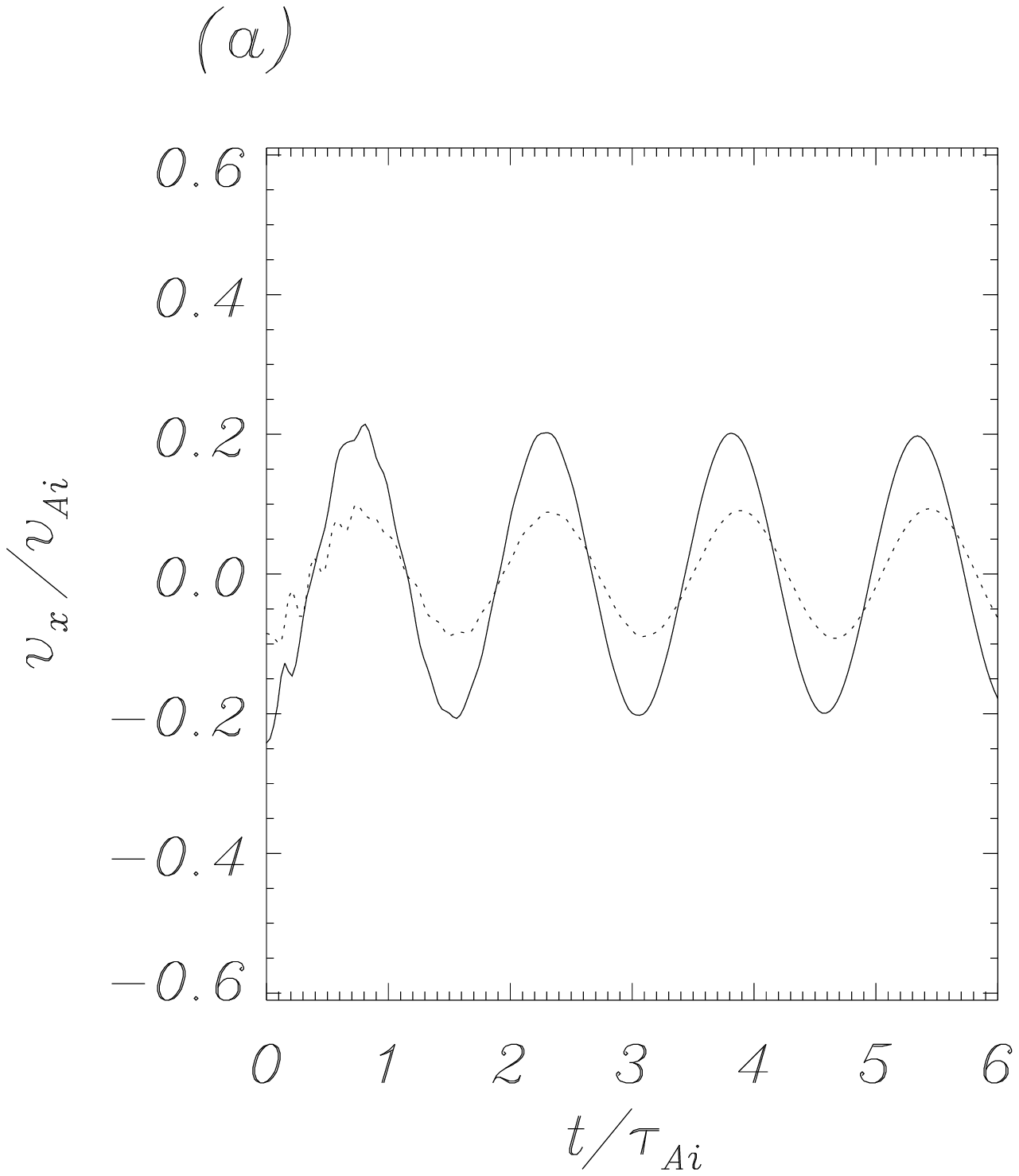}\includegraphics[width=4.5cm]{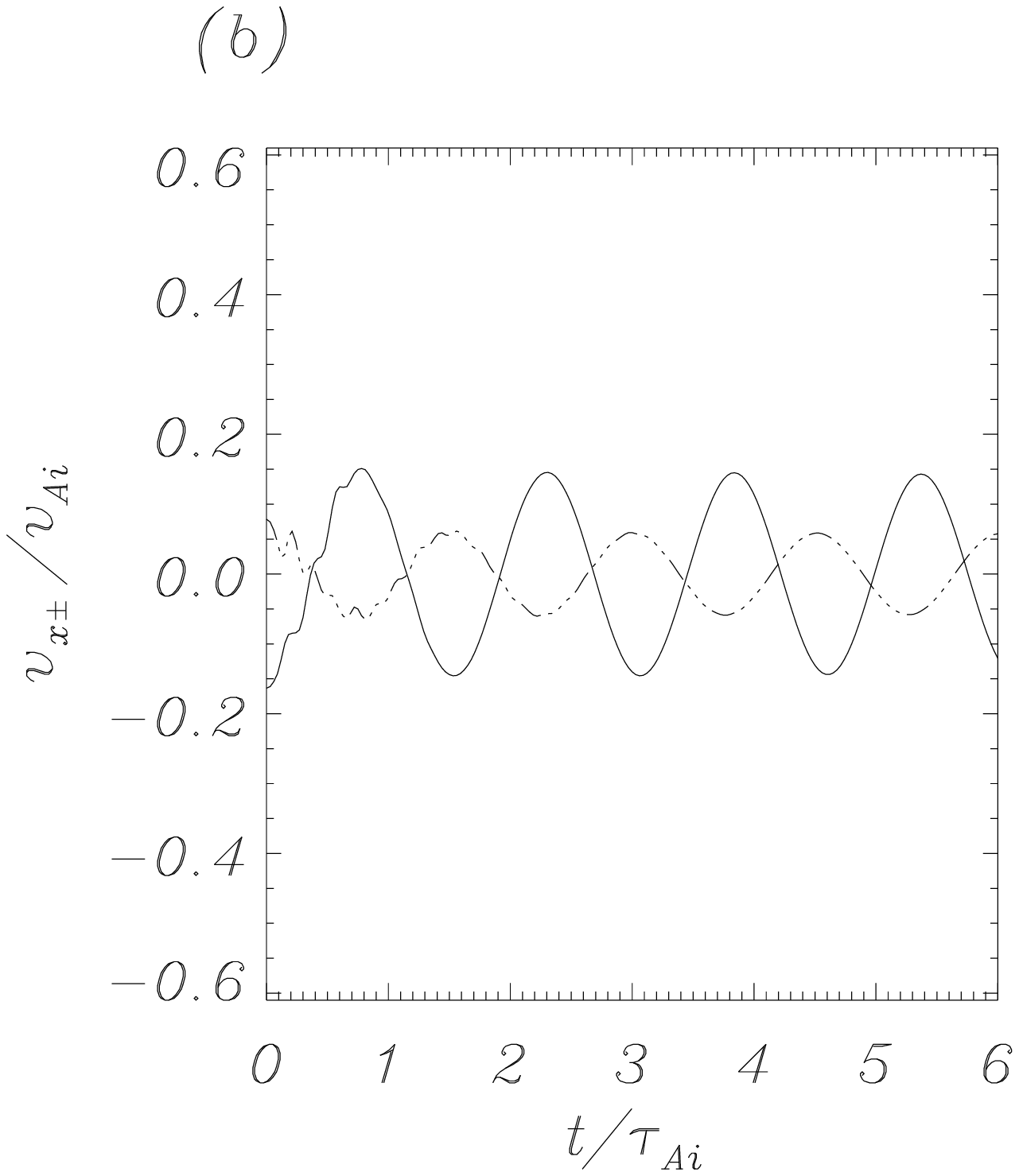}\includegraphics[width=4.5cm]{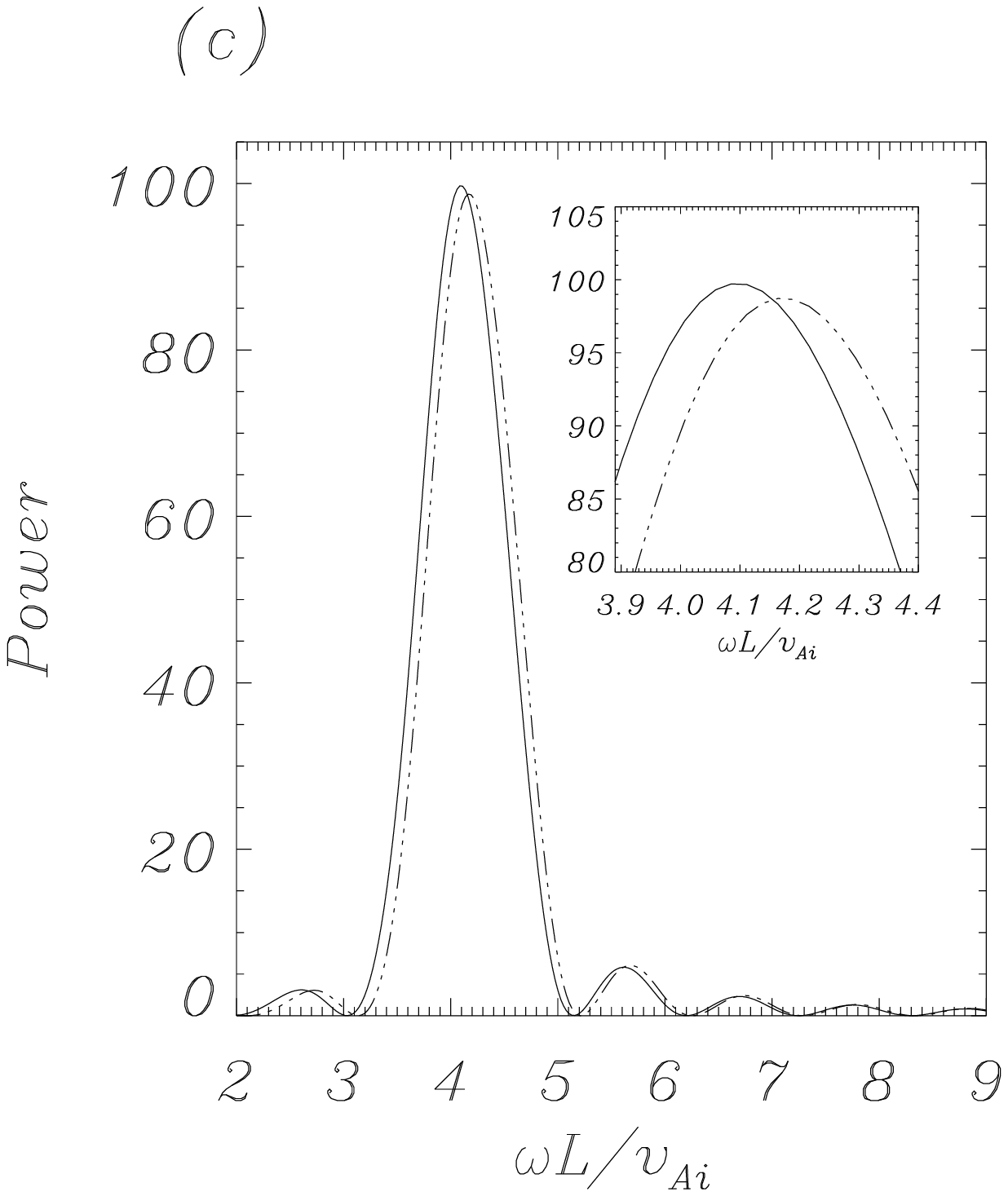}
\mbox{\hspace{-0.cm}\hspace{4.cm}\hspace{4.cm}}\vspace{-0.8cm}
\includegraphics[width=4.5cm]{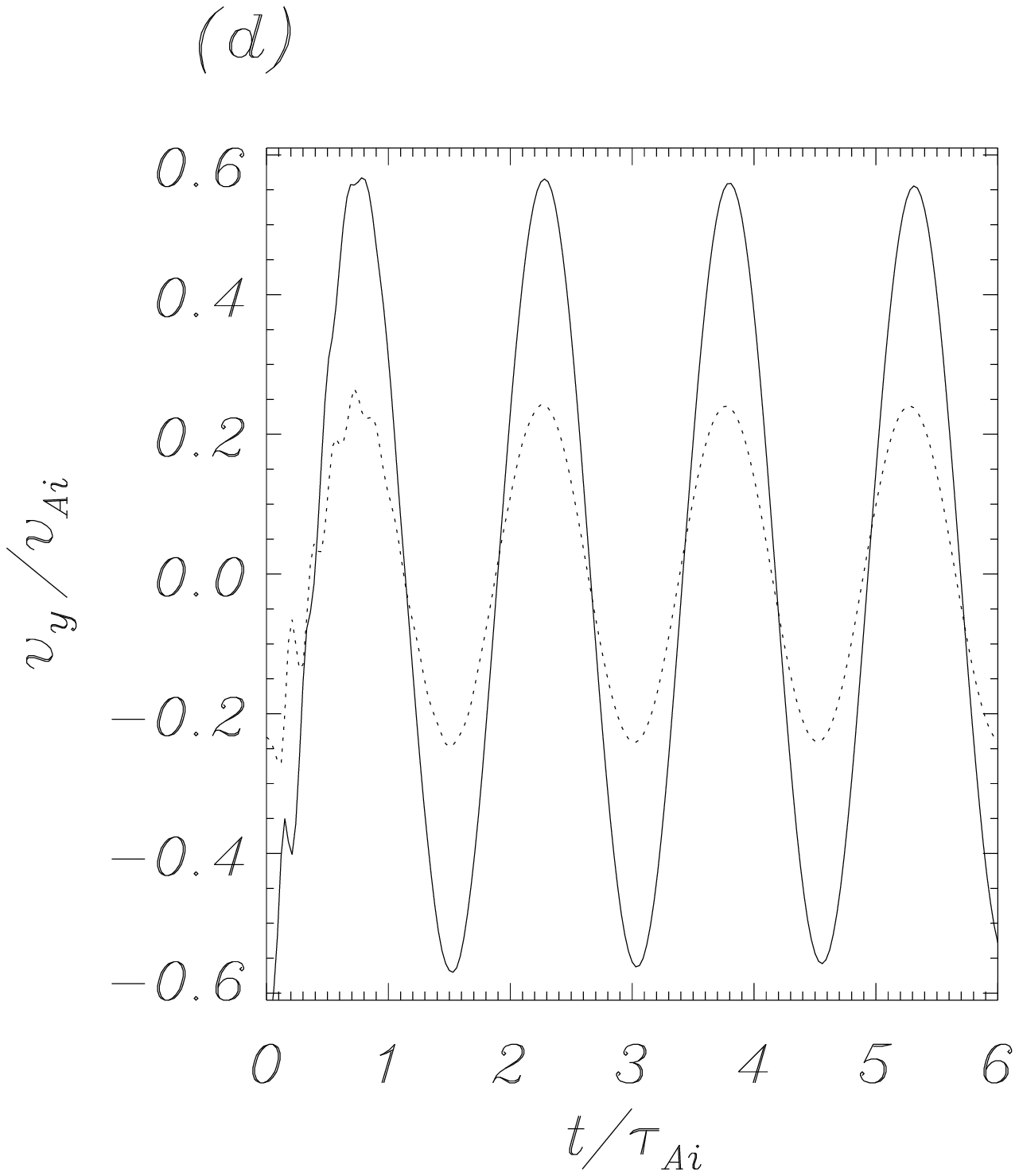}\includegraphics[width=4.5cm]{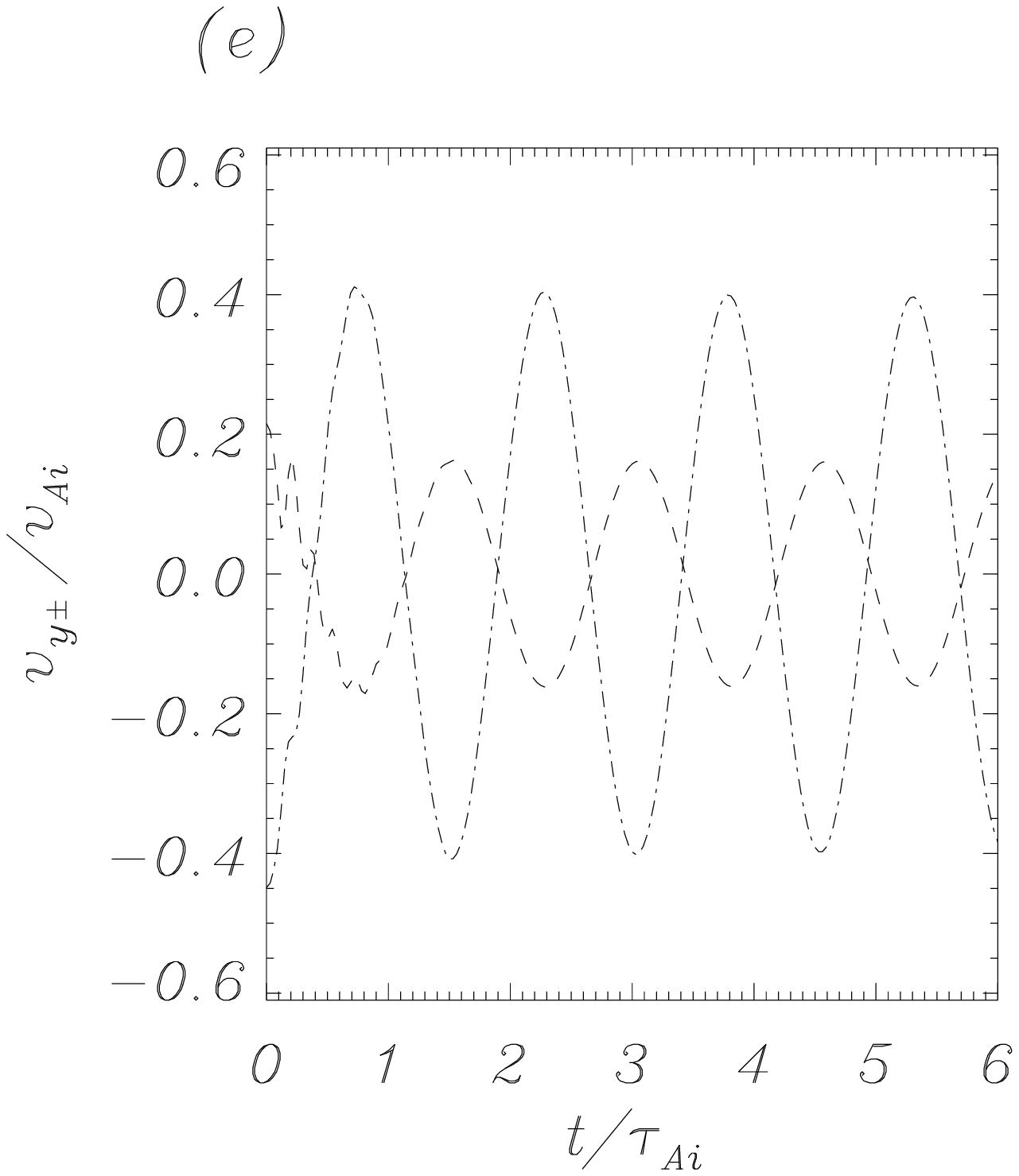}\includegraphics[width=4.5cm]{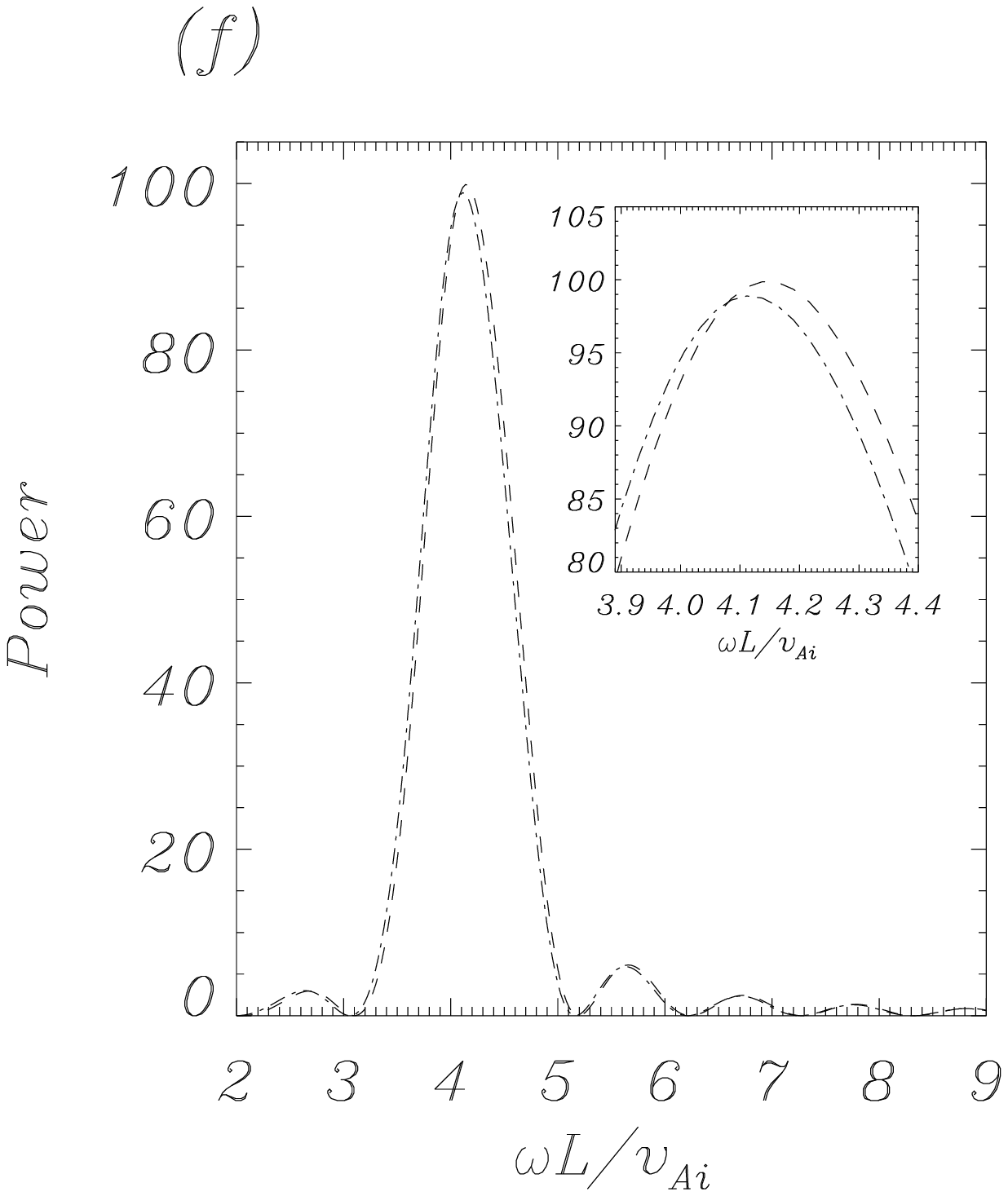}
\caption{
Same as Figure~\ref{time_evol_example_1} for an initial incidence angle
$\alpha=70^\circ$.
}
\label{time_evol_example_2}
\end{figure}
\clearpage

In Figures \ref{time_evol_example_1}a and \ref{time_evol_example_1}d the $x$-
and $y$-components of the velocity at the center of each loop are plotted. In
these figures we see that, after a very brief transient characterized by
short-period oscillations, the system oscillates with the sum of normal modes.
The frequencies of the modes are quite similar, and it is difficult to resolve
them. Although the frequencies of these modes are present in the time-dependent
signal, this information cannot be easily extracted from the data because in
these simulations the maximum evolution time (which is determined by the
numerical damping) is $T=6\tau_\mathrm{Ai}$. With this maximum time we have a
frequency resolution $2/T \simeq 0.3/\tau_\mathrm{Ai}$, but, as evidenced by
Figure~\ref{w_vs_d}, the difference in frequency between the eigenmodes is
typically less than $0.1/\tau_\mathrm{Ai}$ so we have not enough frequency
resolution. For this reason we extract the frequencies with another method
considering that the velocity field is the addition of normal modes with
symmetric and antisymmetric spatial functions with respect to the $y$-axis. We
measure the velocity in the loop centers $(x=-d/2,y=0)$ and $(x=d/2,y=0)$, i.e.
two symmetric points with respect to $x=0$. Then, the sum of both measured
velocities in these points is twice the symmetric velocity. Dividing this
velocity by two we obtain the $v_x$ of the $S_x$ mode and the $v_y$ of the $S_y$
mode in these points. On the other hand, the subtraction of the measured
velocities is twice the antisymmetric velocity. Similarly, dividing this
velocity by two we obtain the $v_x$ of the $A_x$ mode and the $v_y$ of the $A_y$
mode.  The obtained mode velocities are plotted in
Figures~\ref{time_evol_example_1}b and \ref{time_evol_example_1}e. Next, we
compute a periodogram of these signals (plotted in
Figs.~\ref{time_evol_example_1}c and \ref{time_evol_example_1}f), from which the
frequencies of the collective modes are determined. The periodogram is preferred
over the FFT as it allows to precisely identify these frequencies.

The above procedure has been applied to numerical simulations for different
separations between loops and the frequencies of the four fundamental eigenmodes
have been obtained. The calculated frequencies have been superimposed to the
normal mode values in Figure~\ref{w_vs_d} using symbols. A good agreement
between the normal mode calculations and the time-dependent results can be
appreciated.

\subsection{Effect of the incidence angle} \label{study_alpha}

We next study the evolution of the system for different incidence angles,
$\alpha$, of the planar pulse and a fixed distance between loops ($d=6 a$). Some
examples of the time evolution have already been discussed and shown in
Figures~\ref{t_evol_90}, \ref{t_evol_0}, and \ref{t_evol_45}. The mode
excitation depends on the width, $\Delta$, the incidence angle, $\alpha$, and
the position, $\mathbf{r}_\mathrm{0}$, of the initial disturbance, but here we
only consider the dependence on the incidence angle. The angles considered in
our simulations vary from $\alpha=0^\circ$ to $90^\circ$ with steps of
$5^\circ$. Using the method of \S \ref{time_dependent_normal_modes} it is also
possible to extract the amplitude of each normal mode, given by the amplitude of
the sinusoidal oscillations in the stationary phase. Two examples of the
extraction method are plotted in Figure  \ref{time_evol_example_1} for
$\alpha=45^\circ$ and Figure  \ref{time_evol_example_2} for $\alpha=70^\circ$.

\clearpage
\begin{figure}[!th]
\center
\includegraphics[width=10.cm]{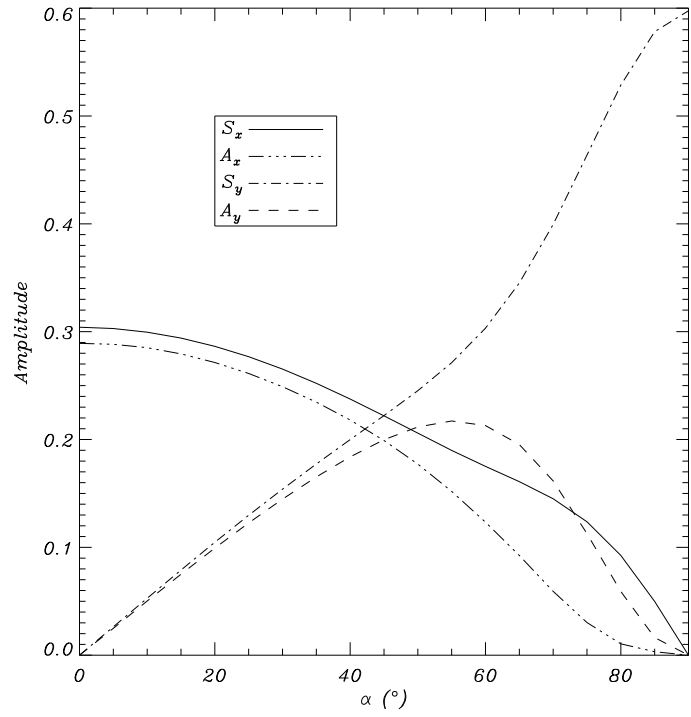}
\caption{
Amplitudes of the four normal modes as a function of the incidence angle
$\alpha$. The separation between loops is $d=6 a$.
} \label{amplitudes}
\end{figure}
\clearpage

In Figure \ref{amplitudes} the amplitude of the four collective modes is plotted
as a function of the incidence angle. The behavior of the amplitude can be
divided in two parts, namely for $0^\circ\le\alpha\le50^\circ$ and for
$50^\circ<\alpha\le90^\circ$. In the first interval the amplitudes of the $S_x$
and $A_x$ modes are more or less equal (see Figs.~\ref{time_evol_example_1}b and
\ref{time_evol_example_1}e as an example) and can be approximated by $0.3 \cos
\alpha$. The same occurs for the amplitudes of the $S_y$ and $A_y$ modes, which
vary roughly as $0.29 \sin \alpha$. In the second interval these amplitudes can
be quite different (see Figs. \ref{time_evol_example_2}b and
\ref{time_evol_example_2}e as an example) and the $S_x$, $A_x$, and $A_y$
amplitudes go to zero at $\alpha=90^\circ$. On the other hand, the $S_y$
amplitude increases and reaches its maximum value at $\alpha=90^\circ$.
Furthermore, for $\alpha=0^\circ$ the amplitudes of the $S_x$ and $A_x$ modes
have a maximum around $0.3$ while the amplitudes of $S_y$ and $A_y$ modes are
zero. This is because for $\alpha=0^\circ$ the initial disturbance drives the
$x$-component of the velocity and so only the $S_x$ and $A_x$ modes are excited.
Similarly, for the perturbation with $\alpha=90^\circ$ only the $S_y$ and $A_y$
modes can be excited, although the shape of our initial perturbation prevents
the $A_y$ mode from being driven and so the $S_y$ mode reaches the largest
amplitude of all modes. On the other hand, the excitation of the antisymmetric
modes requires the initial disturbance to hit the right and left loops at
different times. For this reason, the amplitudes of these modes decrease with
$\alpha$. In fact, when $\alpha=90^\circ$ this time difference is zero since
both loops are excited at the same time and the amplitude of the $A_x$ and $A_y$
vanishes. Finally, it is interesting to note that for $\alpha=45^\circ$ the four
modes are excited with almost the same amplitude.

\section{Study of the loops motions: beating}
\label{2d_beating}

As we have shown in the previous section, loop motions can be very complex. This
is even more clear in Movie 4, in which the time-evolution for a simulation with
identical parameters to those used in Figure~\ref{t_evol_45} but for much larger
times is represented. In \S \ref{time_dependent_analysis} we mentioned that the
initial disturbance excites the right loop but does not perturb the left loop.
After a short time the left tube starts to oscillate due to the interaction with
the right one. At this stage, the right loop oscillates with the velocity
polarization of the initial pulse, whereas the left tube oscillates in a
direction perpendicular to that of the initial disturbance.  The reason for the
complexity of the loops motions is that their oscillations are a superposition
of four normal modes with different velocity polarizations, parities, and
frequencies.

\clearpage
\begin{figure}[!ht]
\center	
\includegraphics[width=8cm]{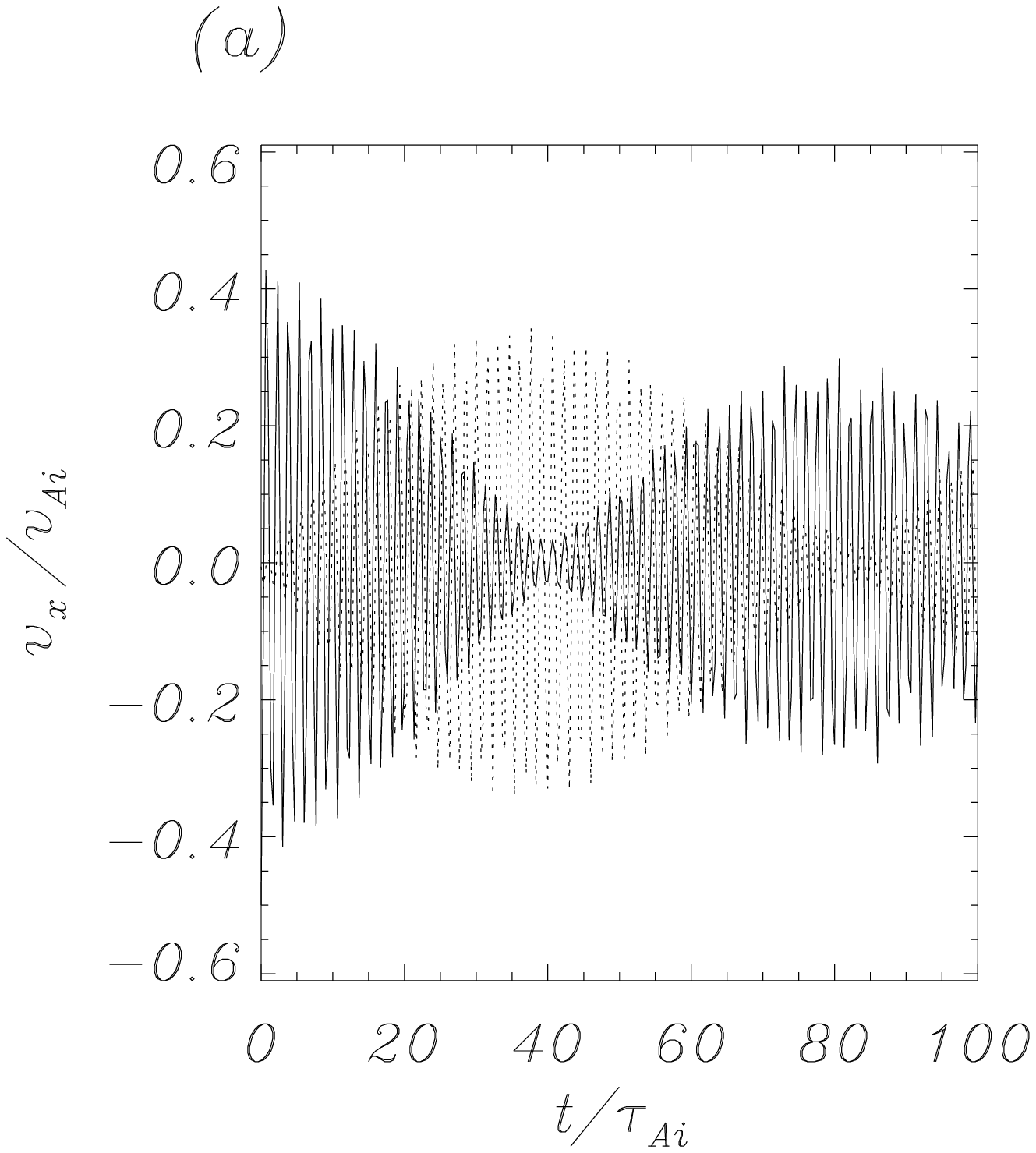}
\includegraphics[width=8cm]{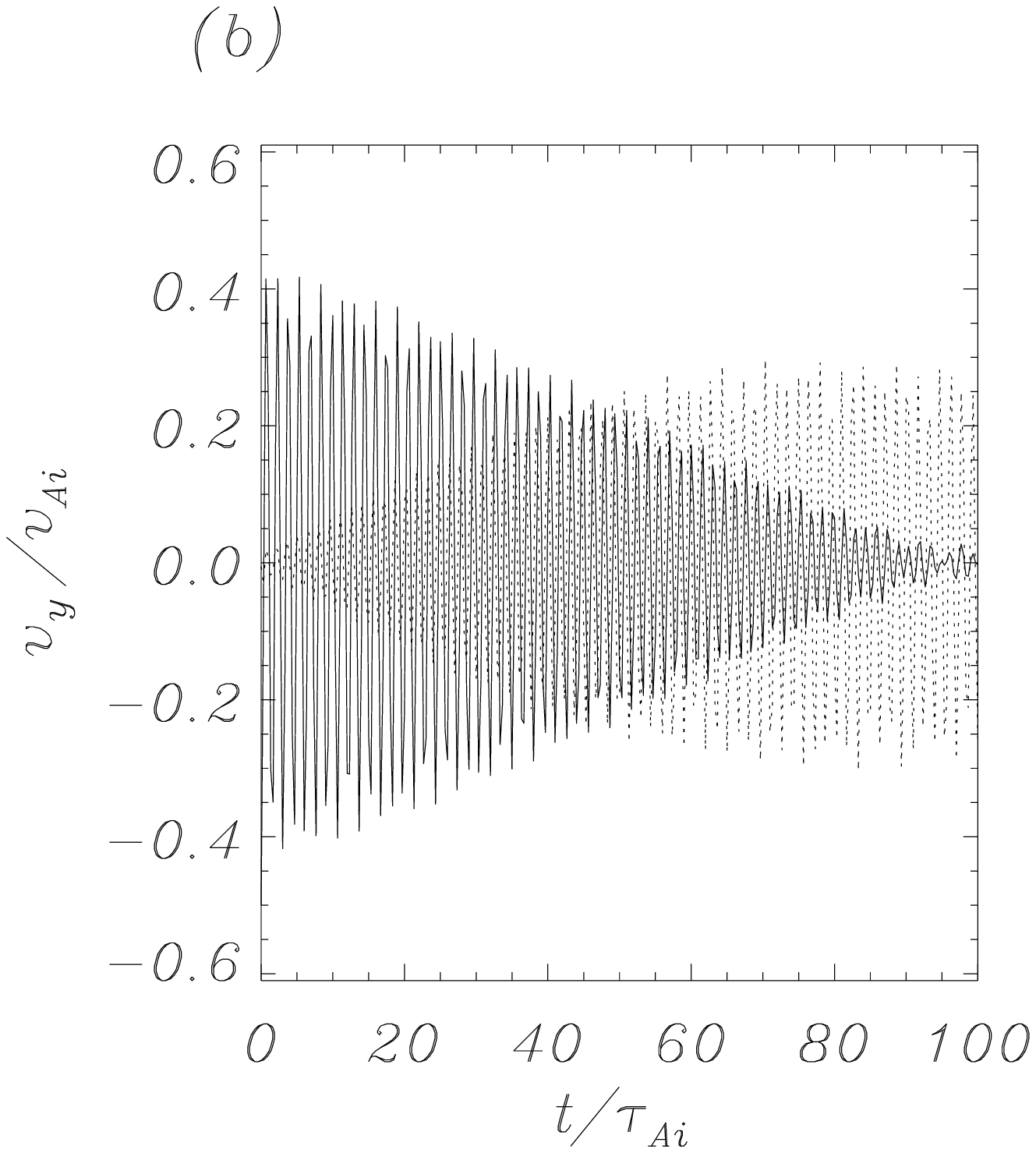}
\caption{
Temporal variation of the velocity components (a) $v_x$ and (b) $v_y$ at the
center of the right loop (solid line) and left loop (dashed line). These results
correspond to the simulation shown in Figure \ref{t_evol_45} and illustrate the
beating of the pair of loops. Damping caused by numerical dissipation causes a
slight decrease of  the amplitude during the numerical simulation. The
time-evolution is also available as an mpeg animation in Movie 4. } 
\label{beating_vxy}
\end{figure}
\clearpage
We next analyze this case in detail. The $x$- and $y$-components of the velocity
at the center of the loops are represented in Figures \ref{beating_vxy}a and
\ref{beating_vxy}b, respectively. There is a clear beating, characterized by the
periodic interchange of the $x$- and $y$-components of the velocity between the
loops. The two velocity components are modulated in such a way that $v_x$
reaches its maximum value in the left tube and becomes zero in the right tube at
the same time (around $t\simeq40 \tau_\mathrm{A i}$). This process is reversed
at $t\simeq80 \tau_{A i}$ and repeats periodically. 

The loops motions can be studied theoretically. In the stationary phase, during
which the system oscillates in the normal modes $S_x$, $A_x$, $S_y$, and $A_y$,
the velocity field components are
\begin{eqnarray}
\label{ss_beat_x}
v_x (x,y,t)&=& C_x^\mathrm{S} (x,y) \cos \left( \omega_x^\mathrm{S} t+
\phi_x^S\right)+C_x^\mathrm{A} (x,y)\cos \left( \omega_x^\mathrm{A} t+
\phi_x^A\right) ,\\ 
\label{ss_beat_y}
v_y (x,y,t)&=& C_y^\mathrm{S} (x,y)\cos \left( \omega_y^\mathrm{S} t+
\phi_y^S\right)+C_y^\mathrm{A} (x,y)\cos \left( \omega_y^\mathrm{A} t+
\phi_y^\mathrm{A}\right).
\end{eqnarray}
The $\mathrm{S}$ and $\mathrm{A}$ superscripts refer to the symmetric and
antisymmetric modes, respectively. The functions $C_x^\mathrm{S}$,
$C_x^\mathrm{A}$, $C_y^\mathrm{S}$, and $C_y^\mathrm{A}$ represent the spatial
distribution of the four normal modes (see Fig.~\ref{normal_modes}) and their
amplitude accounts for the energy deposited by the initial disturbance in each
of them. The normal mode frequencies are represented by their frequencies,
$\omega$, while $\phi$ are their initial phases.

Let us turn our attention to the results in
Figure~\ref{time_evol_example_1}. In the loops centers the symmetric and
antisymmetric modes have a very similar amplitude (see also
Fig.~\ref{beating_vxy} for $\alpha=45^\circ$), which means that $C_x^\mathrm{S}
(d/2,0) = C_x^\mathrm{A} (d/2,0)$. Then, taking into account the parity of
$C_x^\mathrm{S}$ and $C_x^\mathrm{A}$ about $x=0$, we have $C_x^\mathrm{S}
(-d/2,0) = -C_x^\mathrm{A} (-d/2,0)$. Inserting these expressions into
equations~(\ref{ss_beat_x}) and (\ref{ss_beat_y}) evaluated at the loop centers
we obtain 
\begin{eqnarray} 
\label{right_velocity_approax} 
\mathbf{v}_\mathrm{right} (t) &=&\Big(C_x \cos ( \frac{\omega_x^\mathrm{A}-\omega_x^\mathrm{S}}{2}~t
)\cos ( \frac{\omega_x^\mathrm{A}+\omega_x^\mathrm{S}}{2}~t
) , C_y \cos (\frac{\omega_y^\mathrm{A}-\omega_y^\mathrm{S}}{2}~t )\cos
(\frac{\omega_y^\mathrm{A}+\omega_y^\mathrm{S}}{2}~t) \Big) ,\\
\label{left_velocity_approax}
\mathbf{v}_\mathrm{left} (t) &=&-\Big(C_x \sin (\frac{\omega_x^\mathrm{A}-\omega_x^\mathrm{S}}{2}~t
) \sin (\frac{\omega_x^\mathrm{A}+\omega_x^\mathrm{S}}{2}~t
) ,C_y \sin (\frac{\omega_y^\mathrm{A}-\omega_y^\mathrm{S}}{2}~t
) \sin (\frac{\omega_y^\mathrm{A}+\omega_y^\mathrm{S}}{2}~t
) \Big).
\end{eqnarray}
where $\mathbf{v}_\mathrm{right}$ and $\mathbf{v}_\mathrm{left}$ are the
velocity of the right and left loop, respectively. We have defined $C_x=2
C_x^\mathrm{S}(d/2,0)$ and $C_y=2 C_y^\mathrm{S}(d/2,0)$ and have assumed
$\phi_x^\mathrm{S}=\phi_x^\mathrm{A}=\phi_y^\mathrm{S}=\phi_y^\mathrm{A}=0$
because the initial disturbance is over the right loop. The beating curves shown
in Figure~\ref{beating_vxy} are accurately described by these equations.

These formulae contain products of two harmonic functions. Then, the
temporal evolution during the stationary phase is governed by four periods: the
two oscillatory periods,
\begin{eqnarray}
\label{oscillating_period_x}
T_x={ 4 \pi \over \omega_x^\mathrm{A}+\omega_x^\mathrm{S}} ,\\
\label{oscillating_period_y}
T_y={ 4 \pi \over  \omega_y^\mathrm{A}+\omega_y^\mathrm{S}} ,
\end{eqnarray}
{giving the mean periods of the time signal;} and two beating periods,
\begin{eqnarray}
\label{beating_period_x}
T_{bx}={ 4 \pi \over  \omega_x^\mathrm{A}-\omega_x^\mathrm{S}} ,\\
\label{beating_period_y}
T_{by}={ 4 \pi \over  \omega_y^\mathrm{A}-\omega_y^\mathrm{S}}.
\end{eqnarray}
giving the periods of the envelop of the time signal. To apply these expressions
to the numerical simulation of Figure~\ref{time_evol_example_1} we insert the
values of $\omega_x^\mathrm{S}$, $\omega_x^\mathrm{A}$, $\omega_y^\mathrm{S}$,
and $\omega_y^\mathrm{A}$ for $d=6 a$ into
equations~(\ref{oscillating_period_x})--(\ref{beating_period_y}). Then we obtain
$T_x=1.52\tau_\mathrm{A i}$, $T_y=1.52\tau_\mathrm{A i}$,
$T_{\mathrm{b}x}=159.96\tau_\mathrm{A i}$, and $T_{\mathrm{b}y}=479.88
\tau_\mathrm{A i}$. The two oscillating periods are equal because the frequency
distribution is approximately symmetric around the central value (the kink
frequency of an individual loop) for sufficiently large distances (see Fig.
\ref{w_vs_d}). The two beating periods derived from the numerical simulations
match very well these values because Figure~\ref{beating_vxy} gives
$T_{\mathrm{b}x}/4\simeq 40 \tau_\mathrm{A i}$ and $T_{\mathrm{b}y}/4\simeq 120
\tau_\mathrm{A i}$.

The $\pi/2$ phase difference between $\mathbf{v}_\mathrm{right}$ and
$\mathbf{v}_\mathrm{left}$ (see Figs.~\ref{time_evol_example_1}a and
\ref{time_evol_example_1}d) is due to the fact that our system of two loops
basically behaves as a pair of driven-forced oscillators. Considering $v_x$, the
left loop has initially a $\pi/2$ delay with respect to the right loop because
it behaves as a driven oscillator and the left one like a forced oscillator.
After half beating period, $T_\mathrm{b x}/2$, the roles are exchanged and left
loop becomes the driver and right one the forced oscillator. The $y$-components
of $\mathbf{v}_\mathrm{right}$ and $\mathbf{v}_\mathrm{left}$ exhibit the same
behavior (see Fig.~\ref{time_evol_example_1}d). This  was already shown by
\citet{luna} in the case of two slabs.

As we have seen, the polarization of the oscillations changes with time (see
Movie 4 for an example). In the beating range, we can see this from the
equations by calculating the scalar product of the velocity at the loop centers,
\begin{eqnarray}\nonumber{}
\mathbf{v}_\mathrm{right} \cdot \mathbf{v}_\mathrm{left}= -C_x^2 \sin \Big(2
(\omega_x^\mathrm{A}-\omega_x^\mathrm{S}) t\Big)\sin \Big(2 (\omega_x^\mathrm{A}+\omega_x^\mathrm{S})t\Big)\\ 
-C_y^2 \sin \Big(2
(\omega_y^\mathrm{A}-\omega_y^\mathrm{S})t\Big)\sin \Big(2 (\omega_y^\mathrm{A}+\omega_y^\mathrm{S})t\Big) .
\end{eqnarray}
This product gives the relative polarization of the loop oscillations and we see
that is zero at $t=0$ and approximately zero for sufficiently small times. Thus,
the left loop does not oscillate initially and it starts to oscillate
perpendicularly to the right loop during the first oscillations. This feature is
shown in Figure \ref{t_evol_45} and Movie~3 and Movie~4.

Similar beating features are recovered for incidence angles of the initial
disturbance in the range $0^\circ \le \alpha \lesssim 50^\circ$ (what we call
the beating range). The cause of this behavior is explained by
Figure~\ref{amplitudes}: for these values of $\alpha$ a similar amount of energy
is deposited in the $S_x$ and $A_x$ modes, so the beating of the $v_x$ component
is possible. Obviously, an analogous argument applies to $v_y$. This is not the
case for $50^\circ \lesssim \alpha \le 90^\circ$ for which the symmetric and
antisymmetric modes receive different amounts of energy from the initial
excitation and then their relative amplitude is different (see Fig.
\ref{time_evol_example_2} for an example). Simulations for angles 
$\alpha>50^\circ$ do not clearly exhibit beating and the trajectories of the
loops are much more complex than those in the beating range.

\section{Discussion and conclusions}
\label{discussion_conclusions}

In this work, we have investigated the transverse oscillations of a system of
two coronal loops. We have considered the low-$\beta$, ideal MHD equations and
have studied both the normal modes of this configuration and the time-dependent
problem. The results of this work can be summarized as follows:

\begin{itemize}

\item The system has four fundamental normal modes, somehow similar to the kink
mode of a single cylinder. These modes are collective, i.e. the system
oscillates with a unique frequency, different for each mode. When arranged in
increasing frequency the modes, are $S_x$, $A_y$, $S_y$, and $A_x$, where
$S$($A$) stand for symmetric (antisymmetric) velocity oscillations with respect
to the plane in the middle of the two loops and $x$ ($y$) stands for the
polarization of motions. These modes produce transverse motions of the tubes, so
they are kink-like modes.

\item We have studied the eigenfrequencies as a function of the separation of
loops. For large distances between cylinders, they behave as a two independent
loops, i.e. the frequency tends to the individual kink mode frequency. When the
distance decreases the frequency splits in four branches, two of which
correspond to the $S_x$ and the $A_y$ modes and are below the frequency of the
individual tube, and the other two are related to the $S_y$ and $A_x$ modes and
lie above the kink frequency of a single tube. Roughly speaking, there is a
certain parallelism between our system of two loops and a mechanical system of
two coupled oscillators with $f$ degrees of freedom, which has $f\times n$
collective normal modes. This parallelism is possible because a slab or a
cylinder oscillating with the kink mode moves more or less like a solid body.
The number of translational degrees of freedom one for an individual slab
($f=1$) and two for an individual loop. Then, the parallelism with the
mechanical system tells us that in a two slab system there are two collective
normal modes \citep{luna}, while in a two cylinder system there are four.

\item For small distances between the loops, the frequency of the $S_x$ and
$A_y$ modes is quite similar and tends to the internal cut-off frequency. This
is different to the behavior in a configuration of two slabs \citep[see][]{luna}
where, for small distances between the slabs, the system behaves as an
individual loop of double width. On the other hand, for the two cylinders the
frequency is much lower than that of a loop with double radius.

\item We have also studied the temporal evolution of the system after an initial
planar pulse. We have shown that, depending on the incidence angle, the system
oscillates with a combination of several normal modes. The frequencies of
oscillation calculated  from the numerical simulations agree very well with the
normal mode eigenfrequencies.

\item In the beating range ($0^\circ \le \alpha \lesssim 50^\circ$), the system
beats in the $x$- and $y$-components of the velocity and the left and right
loops are $\pi/2$ out of phase for each velocity component. They behave as a
pair of driven-forced oscillators, with one loop giving energy to the other and
forcing its transverse oscillations. The role of the two loops is interchanged
every half beating period. On the other hand, for perturbations with
$\alpha>50^\circ$ the loops motions are much more complex than those in the
beating range. The phase lag cannot be clearly appreciated and it strongly
depends on the incidence angle of the initial pulse.

\end{itemize}

From this work, we conclude that a loop system clearly shows a collective
behavior, its fundamental normal modes being quite different from the kink mode
of a single loop. These collective normal modes are not a combination of
individual loop modes. This suggests that the observed oscillations reported in
\citet{aschwanden1999, aschwanden2002, schrijver2002, verwichte2004} are in fact
caused by one or a superposition of some collective modes. Moreover, the
antiphase movements reported by \citet{nakariakov1999} can be easily explained
using our model. The same applies to the bounce movement of loops D and E
studied in \citet{verwichte2004}. These motions can be interpreted by assuming
that there is beating between the loops produced by the simultaneous excitation
of the fundamental $S_x$ and $A_x$ modes.

It should be noted that the observations indicate a very rapid damping of
transverse oscillations, such that in a few periods the amplitude of oscillation
of the loops is almost zero. This fast attenuation may hide the beating
produced by the simultaneous excitation of several normal modes of the system.
However, in some situations, for example, for small loop separations and high
density contrast loops, the beating periods decrease. Then, under such
conditions the beating could be detectable in the observation interval. In
any case, the beating is just one particular collective behavior, and there is
always interaction between the individual loops in short time scales (typically
of the order of $2d/v_\mathrm{A e}$). The consequences of this interaction are
the collective normal modes of the system. The presence of the normal modes
could be also clear from a frequency analysis. Unfortunately, due to the
temporal resolution, these observations do not allow us to perform such
analysis, but the frequency extraction method derived in \S
\ref{time_dependent_normal_modes} is suitable to be applied to the
observations.

Finally, in order to have more realistic models additional effects need to be
included. In this work, we have studied two loops with exactly the same density
and radii, so the next step is to analyze the behavior of a system of $n$ loops
with different properties. This study could also be extended to understand
the possible effect of internal structure (multi-stranded models and small
filling factors) on the oscillating loops by considering a set of very thin
tubes with different densities and radii. We expect that the dynamical behavior
and frequencies of multi-stranded loops differ from those of the monolithic
models. Preliminary work has been done by \citet{arregui2007} who have
studied the effects on the dynamics of the possibly unresolved internal
structure of a coronal loop composed of two very close, parallel, identical
coronal slabs in Cartesian geometry.

\acknowledgments 
M. Luna is grateful to the Spanish Ministry of Science and Education for an FPI
fellowship, which is partially supported by the European Social Fund. He also
thanks the members of the Departament of Mathematics of K. U. Leuven for their
warm hospitality during his brief stay at this University and for their worthy
comments. J. Terradas thanks the Spanish Ministry of Science and Education for
the funding provided under a Juan de la Cierva fellowship. The authors
acknowledge the Spanish Ministry of  Science and Education and the Conselleria
d'Economia, Hisenda i Innovaci\'o of the Goverment of the Balearic Islands for
the funding provided under grants AYA2006-07637, PRIB-2004-10145 and
PCTIB-2005-GC3-03, respectively. We are grateful to the referee for his/her
comments and suggestions that helped to improve the manuscript.

\end{document}